\documentclass[twocolumn]{aastex63}
\usepackage{graphicx}
\graphicspath{{figure/}}
\usepackage{subfigure}
\usepackage{color, hyperref, epsfig}
\usepackage{apjfonts, natbib}
\usepackage{appendix}
\usepackage{amsmath}
\usepackage{float}
\usepackage{bm}
\usepackage{booktabs}
\usepackage{multirow}
\usepackage{ragged2e}
\usepackage{lineno}

\newcommand{\lum}{erg\,s\ensuremath{^{-1}}}

\newcommand{\tdust}{\ensuremath{T\mathrm{_{dust}}}}

\newcommand{\lbol}{\ensuremath{L\mathrm{_{bol}}}}

\newcommand{\qabs}{\ensuremath{Q\mathrm{_{abs}}}}
\newcommand{\msun}{\ensuremath{M_{\odot}}}

\newcommand{\mbh}{\ensuremath{M_\mathrm{BH}}}

\shorttitle{AT 2019qiz DUST MODEL}
\shortauthors{Wu et al.}

\begin{document}

\title{\Large A Torus Remnant Revealed by the Infrared Echo of Tidal Disruption Event AT 2019qiz: Implications for the Missing Energy and Quasiperiodic Eruption Formation}

\author[0009-0006-8112-0187]{Mingxin Wu}
\affiliation{Department of Astronomy, University of Science and Technology of China, Hefei, 230026, China; wmx112611226@mail.ustc.edu.cn, jnac@ustc.edu.cn}
\affiliation{School of Astronomy and Space Sciences,
University of Science and Technology of China, Hefei, 230026, China}

\author[0000-0002-7152-3621]{Ning Jiang}
\affiliation{Department of Astronomy, University of Science and Technology of China, Hefei, 230026, China; wmx112611226@mail.ustc.edu.cn, jnac@ustc.edu.cn}
\affiliation{School of Astronomy and Space Sciences,
University of Science and Technology of China, Hefei, 230026, China}

\author[0000-0003-3824-9496]{Jiazheng Zhu}
\affiliation{Department of Astronomy, University of Science and Technology of China, Hefei, 230026, China; wmx112611226@mail.ustc.edu.cn, jnac@ustc.edu.cn}
\affiliation{School of Astronomy and Space Sciences,
University of Science and Technology of China, Hefei, 230026, China}

\author[0009-0007-1153-8112]{Di Luo}
\affiliation{Department of Astronomy, University of Science and Technology of China, Hefei, 230026, China; wmx112611226@mail.ustc.edu.cn, jnac@ustc.edu.cn}
\affiliation{School of Physical Sciences, University of Science and Technology of China, Hefei, 230026, China}

\author[0000-0002-4757-8622]{Liming Dou}
\affiliation{Department of Astronomy, Guangzhou University, Guangzhou 510006, China} 

\author[0000-0002-1517-6792]{Tinggui Wang}
\affiliation{Department of Astronomy, University of Science and Technology of China, Hefei, 230026, China; wmx112611226@mail.ustc.edu.cn, jnac@ustc.edu.cn}
\affiliation{School of Astronomy and Space Sciences,
University of Science and Technology of China, Hefei, 230026, China}


\begin{abstract}

AT 2019qiz is the first standard optical tidal disruption event (TDE) with detection of X-ray quasi-periodic eruptions (QPEs), providing strong evidence for TDE-QPE association. Moreover, it belongs to the rare subset of optical TDEs with prominent infrared (IR) echoes revealed by the multi-epoch photometry from the Wide-field Infrared Survey Explorer (WISE). The IR light curve shows an early bump, followed by a steady rise until the second-to-last epoch, after which it appears to enter a plateau phase. The dust temperature decreased until the fourth epoch and remains approximately constant for the subsequent five epochs. We have fitted the last five epochs using a convex dust ring model, resulting in an inner radius $>1.2$~pc. Such a large radius greatly exceeds the inner radius of the active galactic nuclei (AGN)  torus for a $10^6$~\msun\ black hole and thus could be a torus remnant with the inner part having vanished, further supporting the unified scenario of recently faded AGNs, TDEs, and QPEs. Consequently, a connection between QPEs and IR-bright TDEs is naturally expected. Moreover, the echo requires at least a peak bolometric luminosity of $(6.6, 9.5, 1.0)\times10^{44}$~\lum\, assuming silicate, silicon carbide, and graphite dust grains, respectively, all of which are significantly higher than the peak optical blackbody luminosity. It adds to the accumulating evidence that the missing energy of TDEs may lie in the unobservable extreme UV. This work highlights the unique value of IR echoes in the study of TDEs and QPEs, and a promising prospect in the era of the Near-Earth Object (NEO) Surveyor, the successor to WISE.

\end{abstract}

\keywords{Infrared astronomy(786); High energy astrophysics (739); Tidal disruption (1696); Time domain astronomy (2109)}

\section{Introduction}

Over the past decade, the detections of transients linked to supermassive black holes (SMBHs) at the centers of galaxies have experienced an explosive growth driven by the advance of multiwavelength time-domain surveys. Among them, the blooming optical surveys dominate the discovery of stellar tidal disruption events (TDEs, see review by \citealt{Gezari2021}), which occur when a star passes close enough to SMBHs that the tidal forces exceed the star’s self-gravity, tearing it apart~\citep{Rees1988}. On the other hand, the X-ray surveys and the targeted X-ray observations of TDE (candidates) have unveiled another population of intriguing transient phenomena called quasi-periodic eruptions (QPEs; e.g., \citealt{Miniutti2019GSN069,Arcodia2021}). In particular, the recent direct detection of QPEs following a standard optical TDE AT 2019qiz~\citep{Nicholl2024Natur} demonstrates that the association with TDE is real for at least some (if not all) QPEs. Additionally, optical TDE hosts and QPE hosts are found to be similar in their morphology and in the overrepresentation of poststarburst galaxies~\citep{Gilbert2024} while the TDE host preference has been alleviated largely by the discovery of a population of infrared (IR)-selected TDEs in starforming galaxies~\cite[]{Jiang2021a,Wang2022,Masterson2024}. Moreover, both hosts show a preference for extending emission line regions~\citep{Wevers2022,Wevers2024-QPE,Wevers2024-TDE}, which are indicative of recently faded active galactic nuclei (AGNs). In the framework of “EMRI+TDE=QPE" model~\citep{Linial2023}, a large fraction of QPE EMRIs are inferred to be quasi-circular, which are likely formed during a previous AGN phase \citep{Zhou2024a, Zhou2024b, Zhou2025}. Based on these facts, a unified scenario is proposed that QPEs are produced after TDEs involving SMBHs when the AGN activity has recently ceased~\citep{Jiang2025embers}. 

\begin{figure*}
\centering
\begin{minipage}{0.9\textwidth}
\centering{\includegraphics[width=1\textwidth]{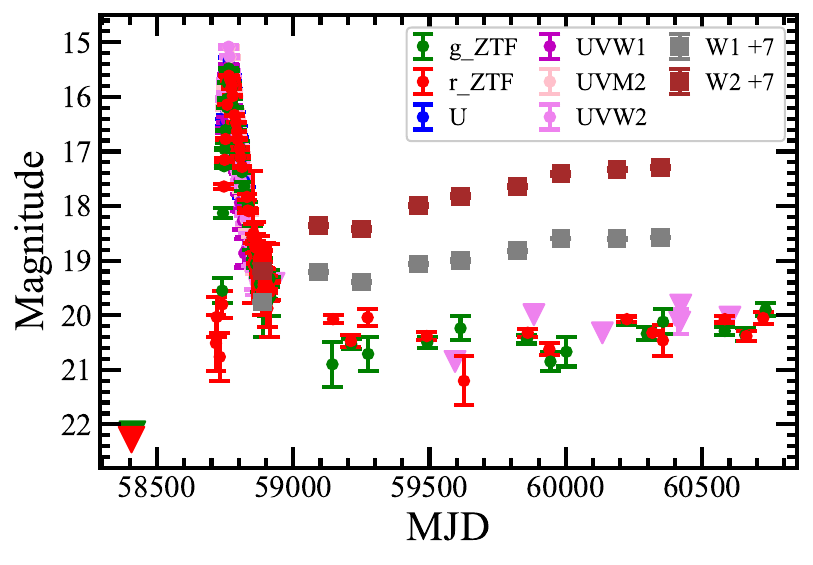}}
\end{minipage}
\caption{The multiwavelength light curve of AT 2019qiz. The Swift/UVOT and ZTF photometry is denoted by dots in different colors, and the WISE photometry by squares. All magnitudes are in the AB magnitude system, except for WISE, which uses the Vega magnitude system.}
\label{lcs}
\end{figure*}

As the first optical TDE with unambiguous detection of QPEs, AT 2019qiz shows a distinctly strong IR echo~\citep{Short2023}, which represents the reprocessed emission of dust in the vicinity of SMBHs~\citep{Lu2016,Jiang2016,vV2016}. A sample study suggests that the IR echoes of vast majority of optical TDEs are actually IR weak, likely due to a very small dust covering factor ($\sim1\%$ or less) and a selection bias of optical surveys~\citep{Jiang2021}. In contrast, the echoes for TDEs in AGNs are much more intense due to the presence of a dusty torus~\citep{Jiang2019}, while it is noteworthy that the population of IR-selected TDEs in inactive galaxies also show comparably strong IR echoes~\citep{Jiang2021a,Wang2022,Reynolds2022,Masterson2024}, indicating a dust-rich subparsec environment. The exploration of the IR echo of AT 2019qiz is of particular interest since it occurs in a recently faded AGN, as evidenced by the detection of an EELR with a scale of $\sim3.7$~kpc~\citep{Xiong2025}, which can provide us with valuable insights into the evolution of the AGN torus. Moreover, the dust IR echo can even help to unveil the intrinsic luminosity and thus the total energy of TDEs, by measuring the reprocessed emission of photons released by TDEs, including the unobserved extreme ultraviolet (EUV) photons, thus opening a new avenue for solving the missing energy puzzle~\citep{Lu2018}. The potential of the IR echo as a bolometer of TDE luminosity has been successfully applied in the study of the extraordinarily long-lived IR echo of PS16dtm, one of the earliest proposed TDEs in AGNs~\citep{Jiang2025PS16dtm}.
However, its application to normal TDEs has been hampered by their generally weak IR echoes, except for a few objects such as AT 2019qiz.

In this work, we present a comprehensive analysis of the remarkable IR echo observed in AT 2019qiz. The echo exhibits an unusually prolonged rising phase followed by a plateau. To interpret this unique light curve, we construct a thin dust torus model with an inner radius of approximately 1 pc and explore key physical properties of the system. Furthermore, we attempt to infer the intrinsic bolometric luminosity of AT 2019qiz and provide indirect evidence for significant missing energy. Finally, we discuss the possible origin of the large-scale torus and the connections between active galactic nucleus (AGNs), IR-bright TDEs, and quasi-periodic eruptions (QPEs).
We assume a cosmology with $H_{0} =70$ km~s$^{-1}$~Mpc$^{-1}$, $\Omega_{m} = 0.3$, and $\Omega_{\Lambda} = 0.7$.

\section{Data}

\subsection{Optical and UV light curves}


All archival data of the Ultraviolet/Optical Telescope \citep[UVOT;][]{Roming2005} onboard the Neil Gehrels Swift Observatory (hereafter Swift) were retrieved and processed using the standard \texttt{Heasoft} pipeline. The images were summed using the \texttt{uvotimsum} task, and light curves were generated with \texttt{uvotsource}, adopting circular source and background regions with radii of $5^{\prime\prime}$ and $40^{\prime\prime}$, respectively. We also obtained the forced photometric light curve of AT 2019qiz from the Zwicky Transient Facility \citep[ZTF;][]{Masci2019}. Since ZTF observations began prior to the outburst of AT 2019qiz, the host galaxy flux could be directly subtracted using pre-event reference images. For the Swift/UVOT data, we used the flux measured from the image taken on UT 2024-10-28 to estimate the host galaxy contribution. All optical photometric measurements were converted to the AB magnitude system and corrected for Galactic extinction before being presented in Figure~\ref{lcs}. The OUV light curve exhibits a rapid rise followed by a steep decline, and subsequently enters a plateau phase (see Figure~\ref{lcs}).

\subsection{Mid-IR light curve}
\label{IRLC}

\begin{figure}
\centering
\begin{minipage}{0.5\textwidth}
\centering{\includegraphics[width=1\textwidth]{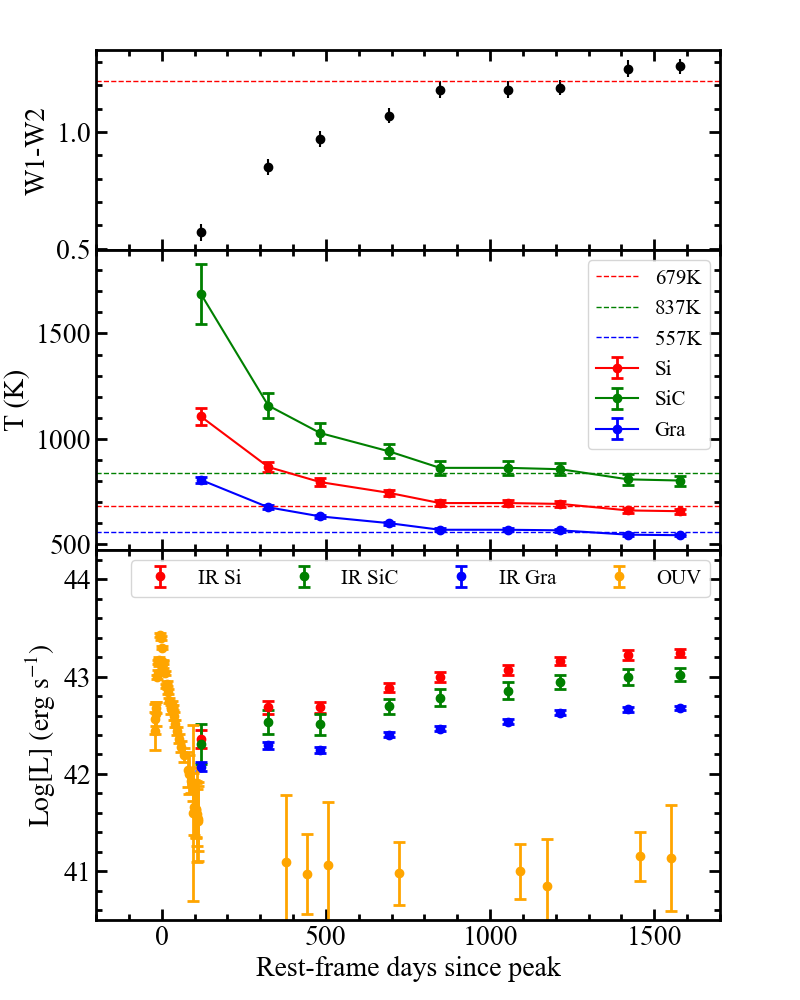}}
\end{minipage}
\caption{Upper panel: evolution of W1-W2. The red dashed line denotes the mean level of the last five epochs. Middle panel: the dust temperature evolution of AT 2019qiz. The red, blue and green full lines denote the temperature of astronomical silicate, graphite and silicon respectively. The dashed lines denote the mean values of the last five epochs. Lower panel: The OUV and IR luminosity evolution of AT 2019qiz. The orange dots denote the OUV blackbody luminosity generated from ZTF and Swift/UVOT data. The red, green, blue dots denote the IR luminosity assuming dust with a radius of 0.1$\mu\text{m}$ composed of Si, SiC and Gra, respectively. The rest-frame days since peak are relative to the date of the optical flare peak (MJD $\sim58767.61$).}
\label{dust}
\end{figure}

The multiepoch mid-infrared (MIR) photometry of AT 2019qiz was obtained from the Wide-field Infrared Survey Explorer \citep[WISE,][]{Wright2010}. The WISE all-sky survey and its extended mission, Near-Earth Object WISE Reactivation~\citep[NEOWISE,][]{Mainzer2014} provide us with a 14-year light curve spanning the period from 2010 February 23 to 2024 February 6, with a half-year cadence, except for a gap between 2010 September and 2014 February. This is the final WISE light curve available, as the mission ended at the end of July 2024. Following the quality control protocols outlined by \cite{Jiang2021}, we implemented rigorous data filtering criteria: excluding observations with poor-quality frame (\texttt{qi\_fact}<1), when the WISE spacecraft was within the boundaries of the South Atlantic Anomaly (\texttt{saa\_sep}<5), scattered moonlight (\texttt{moon\_masked}=1) and artifacts (\texttt{cc\_flags}$\neq$0). The remaining data are then binned every half year to increase the signal-to-noise ratio of the photometry. The WISE light curves are very stable until a sudden brightening on 2020 February 7, which is $\sim120$ days after the optical peak of AT 2019qiz (see Figure~\ref{WISE}). We established the baseline quiescent level by averaging the data before the outburst and then calculated the magnitudes associated with AT 2019qiz with this baseline subtracted. The baseline uncertainty was calculated as the standard error of the mean of the 14 quiescent epochs.
A total of nine epochs of echoes have been detected. There is a bump in the early phase of the IR light curve. After the bump, it had steadily increased until the second to last epoch and then seemed to remain in a plateau phase (see Figure~\ref{lcs}). Compared to the short-lived OUV flare, which decays by $\sim5$ magnitude from its peak within $\sim200$ days, the IR flares last on a much longer time scale, without a declining trend after 4 years since the first detection. 


We computed the properties of dust, specifically focusing on its temperature (\tdust). For simplicity, we assumed a single temperature dust component composed of either astronomical silicate (Si), silicon carbide (SiC), or graphite (Gra), with a grain radius of 0.1$\mu$m. Using the absorption efficiency \qabs($\nu$) from \cite{Laor1993}, the dust temperature is calculated as follows: 
\begin{equation}
\frac{f_{\text{W1}}}{f_{\text{W2}}} = \frac{Q_{\text{abs}}(\nu_{\text{W1}}) B_{\nu}(\nu_{\text{W1}},T_{\text{d}})}{Q_{\text{abs}}(\nu_{\text{W2}}) B_{\nu}(\nu_{\text{W2}},T_{\text{d}})}
\label{Tdust}
\end{equation}
where \(f_{\text{W1}}\) and \(f_{\text{W2}}\) are the flux in W1 and W2 band, \(B_{\nu}(\nu,T)\) is the Planck function, \(T_{\text{d}}\) is dust temperature. Applying this model, we estimate the dust temperature at each epoch. The temperature decreased until the fourth epoch and remained approximately constant throughout the following five epochs (see Figure~\ref{dust}). The dust temperature shows a tentative slight decrease between the seventh and eighth epochs, but it is not significant when considering errors, so we ignore it in our following analysis.
Throughout the entire period of IR detection, the dust temperature remains below the sublimation temperature, except for the first epoch of SiC. We also used the previously derived temperature to estimate the IR luminosity, accounting for the dust absorption efficiency (see Figure~\ref{dust}). It is important to note that these IR luminosities are likely overestimated, as they are calculated assuming dust grains of a single size rather than a realistic size distribution (see Figure~\ref{luminosity_3}). Nevertheless, the estimates still capture the overall trend of IR luminosity evolution: a much longer rising time scale compared to OUV luminosity. This overestimation does not affect the dust model fitting, as our analysis relies on flux in the W1 and W2 bands rather than the derived IR luminosities (see \S\ref{analysis}).

\section{Analysis}
\label{analysis}

\subsection{Dust echo model}

\begin{figure}
\centering
\begin{minipage}{0.5\textwidth}
\centering{\includegraphics[width=0.95\textwidth]{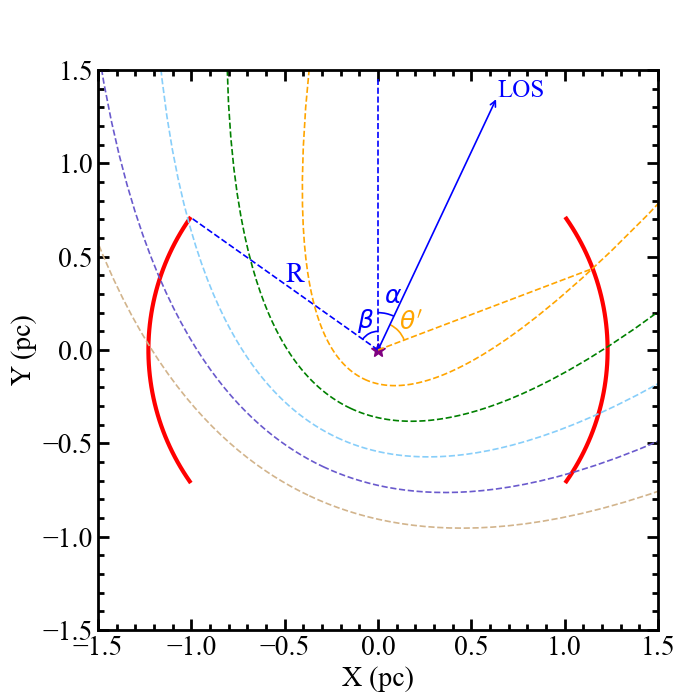}}
\end{minipage}
\caption{Schematic picture of the dust echo model. The central SMBH and dusty torus are shown in purple and red, respectively. The dashed parabolas illustrate the iso-delay surface at lags of 400, 800, 1200, 1600, and 2000 days, respectively.}
\label{model}
\end{figure}

For a spherical shell dust model, the dust located along the line of sight exhibits a minimal time delay, causing the IR light curve to reach the plateau phase too early, on a timescale comparable to the rise of the OUV light curve to its peak. To resolve this inconsistency, we adopt an inclined torus dust model, which avoids small light-travel time delay. Due to the steady increase in IR magnitude, we also infer that the dust is located farther from the central SMBH than the source of the OUV emission, resulting in a significant light-travel time delay. Consequently, the time scale of the OUV light curve is shorter than the light travel time for OUV photons to reach the inner edge of the torus, allowing us to approximate the OUV light curve as a $\delta(t)$ function. Because the inner dust contributes the major IR emission, we simplify the torus to be thin. It is important to note that the dust temperature is not constant during the early phase (see Figure~\ref{dust}). The higher dust temperatures at early times may indicate the presence of additional dust structures located at radii smaller than the inner radius of torus. In this work, we focus on the constant temperature phase, which corresponds to the OUV pulse, and we will discuss the early phase in detail in \S\ref{modelfit}.
Furthermore, given that the dust temperature consistently remains below the sublimation threshold except for the first epoch of SiC, we simply neglect the sublimation process in this model. For simplicity, we assume a single dust composition with a grain radius of 0.1$\mu$m, uniformly distributed along the inner surface of a torus. Based on the above assumptions and simplification, we construct a thin torus dust model similar to that described in \cite{Dou2017} (see Figure~\ref{model}). In this model, the OUV photons should reach all parts of the torus at the same time, but the observed area of dust illuminated by OUV photons from the central source expands over time due to the light time lag(see Figure~\ref{model}). As more heated dust was observed, the IR light curve rises accordingly. The time lag is given by:
\begin{equation}
\tau=t-t_{\text{peak}}=\frac{R}{c}(1-\cos\theta^{\prime})
\label{timelag1}
\end{equation}
or inversely:
\begin{equation}
\theta^{\prime}(t)=\text{arccos}[1-\frac{c}{R}(t-t_{\text{peak}})]
\label{timelag2}
\end{equation}
where \(\tau\) is the time lag, $t$ is the IR observed time, \(t_{\text{peak}}\) is the time of optical flare peak, $R$ is the inner radius of torus, $c$ is light speed, \(\theta^{\prime}\) is the angle between line of sight (LOS) and the direction of OUV light. Because $-1\le\cos\theta^{\prime}\le1$, so we can get the minimum of $R$:

\begin{equation}
R_\text{min}=\frac{c(t_\text{max}-t_\text{peak})}{2}=0.68\text{pc}
\label{Rmin}
\end{equation}
where $t_\text{max}$ corresponds to the time of last epoch.

Then we need to establish two spherical coordinate systems (\(r,\theta,\phi\)) and (\(r^{\prime},\theta^{\prime},\phi^{\prime}\)), with the polar axis of one coinciding with the symmetry axis of the torus, and the other with LOS. So we have: 
\begin{equation}
\theta=\text{arccos}(\cos\alpha  \cos\theta^{\prime}\ - \sin\alpha 
 \sin\theta^{\prime} \cos\varphi^{\prime})
\label{coordinate_tran}
\end{equation}
or inversely:
\begin{equation}
\varphi^{\prime}(\theta,\theta^{\prime})=
\begin{cases}
    \pi &\Delta<-1\\
    \text{arccos}(\Delta) &-1<\Delta<1\\
    0 &\Delta>1
\end{cases}
\label{coordinate_tran2}
\end{equation}
\begin{equation}
\Delta=\frac{\cos\alpha \cos\theta^{\prime}-\cos\theta}{\sin\alpha \sin\theta^{\prime}}
\label{sanjiao}
\end{equation}
where $\alpha$ is the angle between the symmetry axis of the torus and LOS. For a given $t$ and $\Delta t$, using equation~\ref{timelag2}:
\begin{equation}
\Delta\theta^{\prime}=\frac{c\Delta t}{R\sin\theta^{\prime}}
\label{weifen}
\end{equation}
So the area of intersection between the torus and the iso-delay surface is given by:
\begin{equation}
\begin{aligned}
\Delta S(t)&=2[\varphi^{\prime}(\beta,\theta^{\prime}(t))-\varphi^{\prime}(\pi-\beta,\theta^{\prime}(t)]R^2\sin\theta^{\prime}(t)\Delta\theta^{\prime}\\
&=2Rc[\varphi^{\prime}(\beta,\theta^{\prime}(t))-\varphi^{\prime}(\pi-\beta,\theta^{\prime}(t))]\Delta t
\label{area}
\end{aligned}
\end{equation}
where $\beta$ is the half-opening angle of torus. The specific IR flux, \(F_{\nu}(t)\), is given by:
\begin{equation}
\begin{aligned}
F_{\nu}(t)&=\frac{1}{4\pi d^{2}}\sigma_{\text{d}}\Delta S(t)\pi B_{\nu}(\nu,T_d)4\pi a^{2}Q_{\text{abs}}(\nu,a)\\
&=\frac{a^2}{d^2}\sigma_{\text{d}}\Delta S(t)\pi B_{\nu}(\nu,T_d)Q_{\text{abs}}(\nu,a)
\end{aligned}
\label{flux}
\end{equation}
where $a$ is the radius of dust grain, $d$ is the distance from source to the observer, $\sigma_{\text{d}}$ is the surface density, $B_{\nu}(\nu,T)$ is the Planck function. $\Delta t\sim20\,\text{days}$ is the full width at half maxima (FWHM) of the OUV light curve from \cite{Hammerstein2023}. With the dust temperature determined in \S\ref{IRLC}, and using equation~\ref{timelag2},~\ref{area} and~\ref{flux}, the only free parameters of the model are $\sigma_{\text{d}}$, $\alpha$, $\beta$ and $R$.

\subsection{Model fitting}
\label{modelfit}

\begin{figure*}
\centering
\begin{minipage}{1.0\textwidth}
\centering{\includegraphics[width=0.9\textwidth]{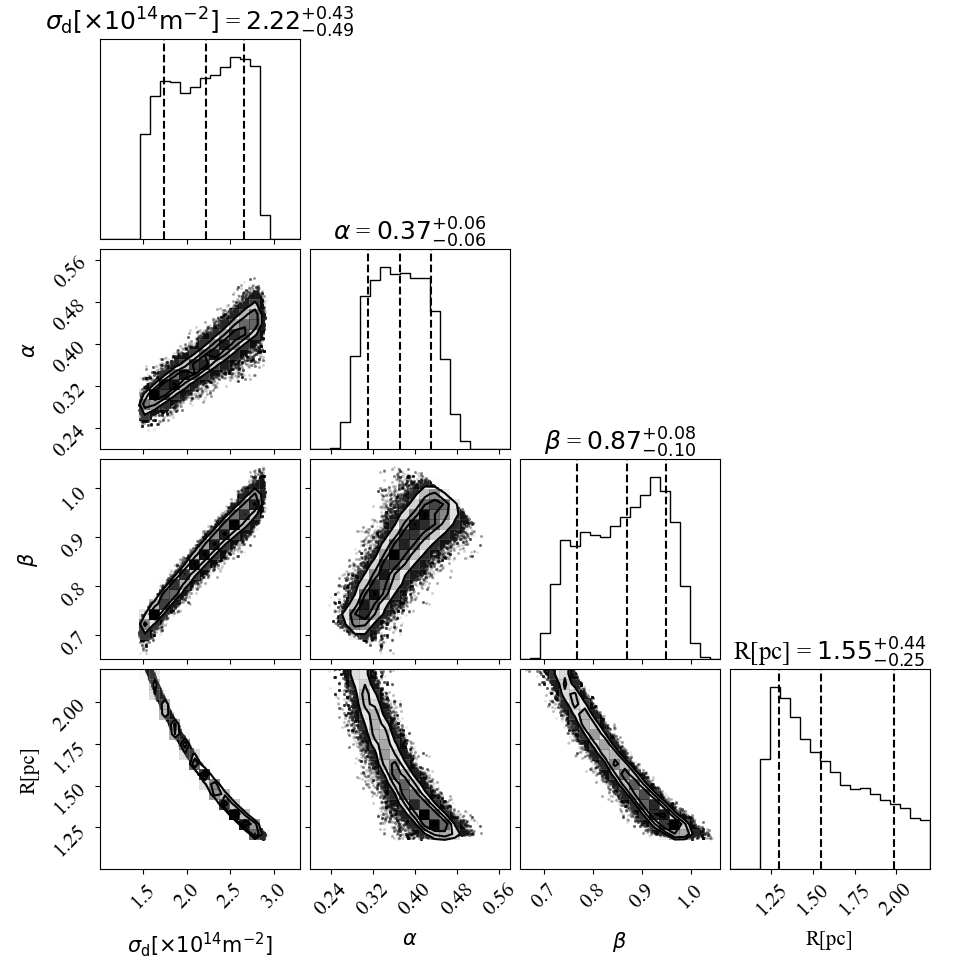}}
\end{minipage}
\caption{The distribution of Si dust echo model parameters for AT 2019qiz. The values and errors above the plots correspond to the $15.87\%$, $50\%$, and $84.13\%$ quantile values of the posterior samples of the parameters, and the black dashed lines show the locations of these values in the plots. The contours in the 2D contour plots reflect the relative numerical density, where darker color means a larger numerical density.}
\label{mcmcsi}
\end{figure*}

\begin{figure}
\centering
\begin{minipage}{0.5\textwidth}
\centering{\includegraphics[width=1\textwidth]{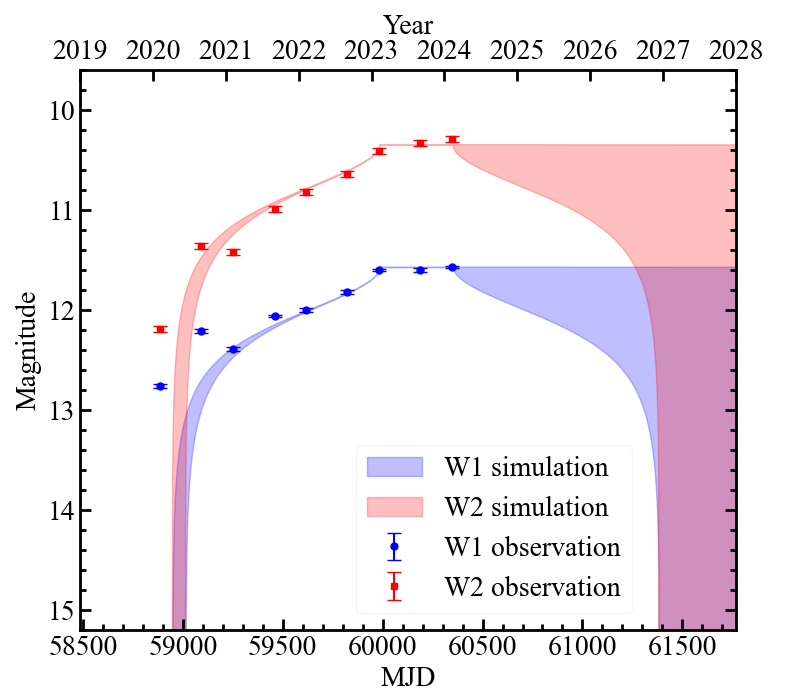}}
\end{minipage}
\caption{Comparison of simulated and observed IR light curves of AT 2019qiz. The blue and red dots denote the observational magnitude of W1 and W2, respectively. The shaded blue and red areas denote the spread of simulated light curve of the top 1$\%$ of posterior samples with the highest likelihood values.}
\label{lightcurve}
\end{figure}

We consider only the last five epochs, during which the dust temperature remains approximately constant. We set a uniform prior for the parameters $\sigma_{\text{d}}$, $\alpha$, $\beta$ and $R$. We apply the Markov Chain Monte Carlo (MCMC) sampler \texttt{emcee}~\footnote{https://emcee.readthedocs.io/en/stable/} \citep{2013PASP..125..306F} to construct the posterior samples of the parameters for dust composed of Si, SiC and Gra, respectively. The prior boundaries of $\alpha$, $\beta$, and $R$ for different dust compositions are set to $[0, \pi/2]$, $[0, \pi/2]$, and $[R_\text{min}, 2\text{pc}]$, respectively. The prior ranges of $\sigma_{\text{d}}$ for Si, SiC, and graphite (Gra) are $[1.5\times10^{14}\,\text{m}^{-2}, 3.5\times10^{14}\,\text{m}^{-2}]$, $[0.4\times10^{14}\,\text{m}^{-2}, 1.8\times10^{14}\,\text{m}^{-2}]$, and $[1\times10^{14}\,\text{m}^{-2}, 3\times10^{14}\,\text{m}^{-2}]$, respectively. The posterior distributions of the three types of dust are shown in Figure~\ref{mcmcsi},~\ref{mcmcsic} and~\ref{mcmcgra} , respectively. The posterior samples are well within the prior boundaries and are sufficiently distant from the edges, indicating that the MCMC sampling is stable and the parameter space is adequately explored. It shows that the distributions of the free parameters except for $\sigma_{\text{d}}$ are almost the same across the three dust compositions, suggesting that the geometrical distribution of dust primarily determines the trend of IR light curve. We select the top 1$\%$ of posterior samples with the highest likelihood values and use them to generate model IR luminosity light curves(see Figure~\ref{lightcurve}). The spread of the light curve during the decline phase is relatively large, because the lack of observational data during the decline phase makes it difficult to constrain the free parameters. Among the parameters $\alpha$, $\beta$, and $R$, all influence the shape of the IR light curve, but observational data are only available in the rising and plateau phases. It is like using two equations to solve for three unknowns: we will get an infinite number of solutions. To illustrate the key role of data in the decline phase of our model fitting, we artificially added a set of data at $\text{MJD}=60700$ with $\text{W1}=11.75\pm0.02$ and $\text{W2}=10.50\pm0.02$, then repeated the fitting for the Si composition. The posterior distributions show that the parameters are significantly better constrained and converge to well-defined values (see Figure~\ref{add}). Meanwhile, the light curve spread during the decline phase has narrowed considerably (see Figure~\ref{lc_add}). Therefore, we expect to obtain much tighter constraints on the parameters when the decline phase is detected in the future.

Our fitting results show that, even among the top 1\% of posterior samples with the highest likelihood values, the parameter space remains relatively broad. Furthermore, likelihood values near the boundaries of the 90\% credible interval are significantly lower than the maximum value, suggesting that some of these samples correspond to poor fits. Therefore, we focus only on samples with high likelihood values. We note that this approach is not used for uncertainty estimation, but rather to explore the region of the parameter space that fits well. Despite the uncertainties, we can still give a lower limitation of the inner radius of the torus, among the top 1$\%$ of posterior samples with the highest likelihood values, $\sim1.2\ \text{pc}$. The values of $\alpha$ and $\beta$ corresponding to the lower limitation of R are the upper limitation, $\sim26^\circ$ and $\sim56^\circ$, respectively. Following the above way, we also tried to fit the model in different sizes and types of dust grains, respectively (see Table~\ref{table}). The results show that $\alpha_\text{max}$, $\beta_\text{max}$, $R_\text{min}$ are almost the same in different sizes and types of dust grains, which are similar to the situation assuming a grain radius of 0.1 $\mu \text{m}$, further suggesting that the trend of the IR light curve is determined by the geometrical distribution of dust, with little dependence on the composition and size of the dust grains.

\begin{table*}[t]
\centering
\caption{Fitting results in different sizes and types of dust grains.} 
\vspace{-7mm}
\label{table}
\begin{center}
\setlength{\tabcolsep}{6mm}{
\begin{tabular}{ccccccccc}
\toprule
Type & a & $T_\text{dust}$ & $\text{log}\,\sigma_\text{d max}$ & $\alpha_\text{max}$ & $\beta_\text{max}$ & $R_\text{min}$ & $\text{log}\,\overline{L_\text{peak}}$ & $\text{log}\,L_\text{peak}$ \\
$\ $ & $\mu\text{m}$ & $\text{K}$ & $\text{m}^{-2}$ & $\ $ & $\ $ & $\text{pc}$ & ~\lum & ~\lum \\
(1) & (2) & (3) & (4) & (5) & (6) & (7) & (8) & (9)\\
\midrule
Si & 0.01 & 683 & 17.4 & $26.2^\circ$ & $56.2^\circ$ & 1.177 &  44.4 &44.6\\
Si & 0.1 & 679 & 14.5 & $26.6^\circ$ & $55.9^\circ$ & 1.174 & 44.7  &44.8\\
Si & 1 & 588 & 11.5 & $26.0^\circ$ & $56.5^\circ$ & 1.175 &  45.3  &45.5\\
SiC & 0.01 & 853 & 17.1 & $26.3^\circ$ & $56.3^\circ$ & 1.174 & 44.2 &44.3\\
SiC & 0.1 & 837 & 14.1 & $26.1^\circ$ & $56.4^\circ$ & 1.175 &  44.8 & 45.0 \\
SiC & 1 & 660 & 10.7 & $26.7^\circ$ & $55.8^\circ$ & 1.175  & 46.0  &46.1\\
Gra & 0.01 & 578 & 17.4 & $26.2^\circ$ & $56.2^\circ$ & 1.178 & 43.4  &43.5\\
Gra & 0.1 & 557 & 14.4 & $26.5^\circ$ & $56.0^\circ$ & 1.174  &  43.8  &44.0\\
Gra & 1 & 729 & 10.5 & $26.0^\circ$ & $56.6^\circ$ & 1.173  &   45.9  &46.1\\
\bottomrule
\end{tabular}
}
\end{center}
\vspace{-1.5mm}
\begin{minipage}{1\textwidth}
\footnotesize
\textbf{Note.} Column (1): Type of dust grains. Column (2): Radius of dust grains. Column (3): The mean temperature of the last five epochs derived in the same way as \S~\ref{IRLC}. The data from column (4) to (7): Upper limitation of the dust surface density, the angle between he symmetry axis of the torus and LOS, the half-opening angle, and the lower limitation of he inner radius of torus, derived in the same way as \S~\ref{modelfit}. The data from column (8) to (9): The pulse bolometric luminosity and the peak bolometric luminosity derived in the same way as \S~\ref{missing-energy}.
\end{minipage}
\end{table*}

We do not include the early-time bump observed in the IR light curve in our model fitting. The simulated IR luminosity is lower than the observed luminosity during this phase, except for the W2 magnitude in the third epoch (see Figure~\ref{lightcurve}). The bump may be generated by structures that are not included in our current model. Since the dust temperature in our analysis remains consistently below the sublimation threshold, we infer that there is few dust at radii smaller than the sublimation radius. One possible explanation is that an additional component in LOS that reprocesses the early OUV photons and re-emits them in the IR, thereby producing the observed bump \citep{Yuan2024}. The additional structure may also cause the higher temperature in the early phase, corresponding to the first four epochs (see Figure~\ref{dust}), if it is in the radius smaller than that of the torus (but still larger than the sublimation radius). Assuming Si dust grain with a radius of 1 $\mu \text{m}$, with the peak bolometric luminosity from \S\ref{missing-energy} and the dust temperature of the first epoch from \S\ref{IRLC}, using equation~\ref{energy-balance}, we can derive that the radius of the dust, which contributes the IR emission of the first epoch, is $\sim0.57\text{pc}$, larger than the sublimation radius and smaller than the inner radius of torus. However, because of the lack of detailed IR spectroscopic data during this phase, we are unable to provide a definitive explanation for the origin of the bump. We note that the W2 magnitude in the third epoch is slightly lower than the simulated value, which is inconsistent with our assumption described above. This discrepancy may be due to our assumption of a pulse-like OUV light curve, whereas the actual light curve is not an idealized pulse shape. This difference may influence the resulting dust temperature and luminosity in early phase. However, the difference is small and does not significantly affect our overall assumptions or conclusions.

\section{Discussion}

\subsection{The Formation of the Large Dust Radius: A Remnant AGN Torus?}
\label{Large-scale-dust-torus}

In this work we have explicitly shown that the continuously rising MIR light curves of AT 2019qiz, which are atypical compared to other optical TDEs~\citep{Jiang2021}, can be well explained by a giant dust structure with a radius of $\sim1.2$~pc. The methodology of this work is entirely inspired by the successful dust echo modeling of another two TDE candidates IRAS~F01004-2237~\citep{Dou2017} and PS16dtm~\citep{Jiang2025PS16dtm}. The commonality among these three events lies in their long slow-rising peak plateau characterized by a relatively steady dust temperature. We note that a completely different interpretation has recently been proposed by \cite{Pasham2025}, who claim that the IR light curves of AT 2019qiz cannot be generated by the TDE itself, including the late-time/remnant TDE disk, but that the reprocessing of the light from the QPEs by a dust shell can reproduce the observations. However, the required peak QPE luminosity is on the order of $\sim10^{44-45}$~\lum, which is one to two orders of magnitude higher than the observed value of $\sim10^{43}$~\lum~\citep{Nicholl2024Natur}. Therefore, the inclusion of QPEs would not yield significant benefits due to their limited energy output. On the other hand, no QPEs were found in PS16dtm, despite it being extensively observed in previous years (e.g. multi-epoch Chandra observations, as discussed in \citealt{Petrushevska2023}). This suggests that QPEs are not the key to producing such unusual, but not unique, IR light curves.


As a standard component of AGNs, the inner radius of the torus is generally thought to be determined by the dust sublimation radius, and thus a $R\propto L_{\rm bol}^{1/2}$ size-luminosity relation is expected and observed (e.g., \citealt{Suganuma2006,Koshida2014}). Using the dust time lags between the WISE and optical light curves of quasars, \citet{Lyu2019} have obtained a similar relationship between the time lags and \lbol\ for quasars i.e.
\begin{equation}
\Delta t_{\rm W1} /\text{day} =10^{2.10\pm 0.06} (\lbol /10^{11} L_{\odot})^{0.47\pm 0.06}
\label{delay1}
\end{equation}
Given a \mbh\ of $\sim10^6$~\msun,  which is the case for AT 2019qiz with \mbh\ estimated from stellar velocity dispersion~\citep{Nicholl2020}, the Eddington luminosity is $\sim10^{44}$~\lum. Thus, the inner radius of the torus will be at most about 0.1~pc if it is simply determined by dust sublimation, even for an Eddington accretion system, either in AGN or TDE phase. Indeed, the dust echoes of optical TDEs and IR-selected TDE candidates all suggest a dust inner radius of $\sim0.1$~pc~\citep{Jiang2021,Jiang2021a,Masterson2024,Necker2025}. Then why is the dust in AT 2019qiz distributed over such a large radius?

One likely possibility is that the torus is the remnant of a recently faded AGN. There are two pieces of evidence. First, the VLT/MUSE observations of the host of AT 2019qiz reveal an EELR ionized by AGN, indicating clearly an AGN phase in the recent past~\citep{Xiong2025}. The other evidence comes from the large dust covering factor ($f_{\text{c}}$) given by our echo modeling. The opening angle $\beta \sim 55^\circ$ corresponds to a $f_{\text{c}} = \cos\beta \approx 0.56$ (e.g. \citealt{Balokovic2018}, see an earlier geometric  definition of $f_c$ in \citealt{Hamann1993}), which is also a typical value of AGNs \citep[e.g.,][]{Stalevski2016}. However, two key questions remain: why does the torus retain a large vertical scale height, and why does the inner dust disappear while the outer dust persists when an AGN fades?  \citet{Krolik1988} proposed that the torus consists of clumps with supersonic random velocities. Frequent collisions between these clumps dissipate orbital energy but boost random motion, which drives dust inward and inflates the torus vertically, respectively \citep[see also in][]{Beckert2004}. If the supply of dust to the outer edge of the torus stops once the AGN has faded, the inner dust becomes denser and falls into the SMBH more quickly due to more frequent collisions, leaving only dust in the outer region. This may also explain the absence of a torus, or more generally a lack of dust, in the vicinity of quiescent SMBHs as revealed by IR echoes~\citep{vV2016,Jiang2021}. Nevertheless, the sustained vertical scale height of the outer torus implies that clump collisions remain frequent even in these regions, suggesting that the structure is dynamically unstable and may eventually dissipate. Following \citet{Krolik1988}, we can crudely estimate the infall velocity of clumps towards the black hole as follows:
\begin{equation}
v_{\text{in}} \sim 4\epsilon_{\text{diff}}pf_{\text{c}}\left(\frac{\Delta v}{v_{\text{orb}}}\right)^3v_{\text{orb}}
\end{equation}
where $\epsilon_{\text{diff}}$ is the degree of the inelasticity, $p$ is fraction of the cloud's mass which participates in the collision, $\Delta v$ is the velocity dispersion of clumps and $v_{\text{orb}}$ is the orbital velocity of clumps. As suggested by \citet{Krolik1988}, collisions between clumps are likely to be highly dissipative, so we assume that $\epsilon_{\text{diff}} \sim1$. We also adopt $p\sim0.2$, consistent with their analysis. For the vertical scale of torus, $\Delta v$ should be comparable to $v_{\text{orb}}$. So we assume that $\frac{\Delta v}{v_{\text{orb}}} \sim 1$, and $v_{\text{orb}}$ is the Keplerian circular orbital velocity, about $70\,\text{km}\,\text{s}^{-1}$ in a radius of 1 pc. So the infall time scale of dust located within a radius of less than 1 pc towards the black hole can be estimated as:
\begin{equation}
t\sim\frac{1\text{pc}}{v_{\text{in}}}\sim3 \times 10^4\, \text{years}
\end{equation}
which also suggests the extinction of AGN activity likely occurred around ten thousand years ago.

We also consider alternative mechanisms for the formation and maintenance of the AGN torus to interpret our simulation results. It has been proposed that radiation pressure is capable of supporting a geometrically thick torus \citep{Krolik2007, Wada2012, Wada2015, Chan2016, Williamson2019}. However, when the AGN fades, the associated radiation pressure decreases, potentially leading to a potential collapse of the torus structure. In the disk wind scenario, the torus is not a static structure, but rather a region within a hydromagnetic outflow composed of clumpy clouds \citep{Emmering1992, Konigl1994, Elitzur2006, Elitzur2009}. This model requires ongoing accretion to sustain the outflow. Once the AGN luminosity drops below a critical threshold, the outflow rate decreases, and the torus structure disappears \citep{Elitzur2006, Elitzur2009}. Other models consider the role of stellar feedback \citep{Wada2002, Hopkins2016} but find that the resulting torus has a limited covering fraction ($\lesssim 0.2$) and a small aspect ratio ($h/R \lesssim 0.3$), respectively. \citet{Schartmann2009} improve upon this by modeling a young nuclear star cluster, where mass and energy injection from stellar winds and supernovae interact to form a filamentary, thick torus-like structure.  In our case, the disappearance of dust within 1~pc may suggest a low supernova rate in this region during the post-AGN phase. This could be due to intense tidal forces inhibiting cloud collapse and suppressing star formation~\citep{Hsieh2021}.
It is worth noting that in the center of our Milky Way, far-IR imaging has uncovered a similar circumnuclear dust ring centered on Sgr $\text{A}^\star$ with a thickness and radius of 0.34~pc and 1.4~pc respectively \citep[][for a similar result]{Lau2013, Latvakoski1999}. This suggests that either the receding pc-scale dust is a fairly common structure in normal galaxies, or that Sgr $\text{A}^\star$ is also a recently faded AGN. 
In summary, the long-term evolution of the torus after AGN activity ceases remains poorly understood due to a lack of observations and comprehensive simulations. In this sense, IR-bright TDEs could be ideal targets for exploring this process, as they are more likely to occur in fading AGNs — a topic that we will discuss in detail in \S\ref{IRTDE-QPE}. 


The host of AT 2019qiz is a poststarburst galaxy and exhibits a recently faded AGN~\citep{Xiong2025}, both of which can greatly enhance the TDE event rate~\citep{Wang2024}. One may wonder if it is possible that a very bright TDE occurred before AT 2019qiz sublimated the inner dust. For a sublimation temperature of 1500K, the required peak bolometric luminosity of a TDE required to clear out the inner 1~pc dust is $\sim10^{46-47}$~\lum~\citep{Tuna2025}. However, such a luminosity is extremely high as it is far beyond the highest known TDEs ($\sim10^{45}$~\lum) according to the luminosity function \citep{Lin2022, Yao2023}. As we will discuss in \S\ref{missing-energy}, so far we cannot completely rule out the possibility that the measured luminosity has been seriously underestimated due to a significant fraction of the missing energy being released in the unobservable extreme UV photons. Indeed, a dust sublimation process and an ultra-high intrinsic bolometric luminosity have been inferred in PS16dtm~\citep{Jiang2025PS16dtm}, while its energy output could be enhanced by the pre-existing AGN disk, which may be true for AT 2019qiz considering a recent AGN phase.

\subsection{Bolometric luminosity and missing energy puzzle}
\label{missing-energy}

In this subsection, we explore the bolometer functionality of IR echoes of AT 2019qiz. We reasonably assume that the dust is optically thin to its own emission, as demonstrated by a recent JWST spectroscopic study of IR-selected TDEs~\citep{Masterson2025}. In addition, the dust is heated and reaches thermal equilibrium on a very short time scale~\citep{Tuna2025},  i.e.
\begin{equation}
 t_{\rm{th}} 
 \approx 3\,\text{s}\,\left(\frac{\kappa}{100\,\text{cm}^2\,\text{g}^{-1}}\right)^{-1}
    \left(\frac{T_d}{10^3\,\rm K}\right)^{-3},
\end{equation}
where $\kappa$ is the opacity and $T_d$ is the dust temperature.
Consequently, the dust instantaneously reprocesses the incident UV/optical transient light into IR emission with negligible transfer time as below.
\begin{equation}
\begin{aligned}
\frac{L_{\text{bol}}}{4\pi R^2}&\int_{0}^{+\infty}\frac{\pi B_{\nu}(\nu,T_\text{OUV})}{\sigma_{\text{T}} T_\text{OUV}^4}\pi a^2 Q_{\text{abs}}(\nu,a) d\nu\\
&=\int_{0}^{+\infty}\pi B_{\nu}(\nu,T_d)4\pi a^2 Q_{\text{abs}}(\nu,a) d\nu
\end{aligned}
\label{energy-balance}
\end{equation}
where $L_{\text{bol}}$ is the bolometric luminosity and $T_\text{OUV}$ is the peak blackbody temperature of the TDE OUV emission.
$\sigma_{\text{T}}$ is the Stefan-Boltzmann constant.

We have simply assumed a pulse-like light curve for the TDE emission in our model, based on the fact that the time scale of the optical emission is much shorter than that of the reprocessed dust emission in the MIR. We can estimate the pulse bolometric luminosity ($\overline{L_{\text{peak}}}$) from Equation~\ref{energy-balance}, that is $4.7\times10^{44}$, $6.8\times10^{44}$ and $7.1\times10^{43}$~\lum\ for dust composed of Si, SiC, and graphite, respectively. However, we know that the real light curve shape of AT 2019qiz is not a pulse yet has a rising time scale of 11.6 days and declining time scale of 17.9 days \citep{Yao2023}. To account for this discrepancy, we assume that the dust emission in the stable temperature phase (last 5 data points in Figure~\ref{dust}) is heated by the average TDE emission during the full width at half maximum (FWHM) period in the light curves. So we can adopt a simple assumption:
\begin{equation}
\int_{\text{FWHM}}L(t)dt=\overline{L_{\text{peak}}}\Delta t
\end{equation}
where $L(t)$ is the evolution of TDE luminosity and $\Delta t$ is the FWHM of the OUV light curve. We use the bolometric luminosity evolution of AT 2019qiz from~\citet{Hammerstein2023} and then derive the peak bolometric luminosity $L_\text{peak}\sim1.4\,\overline{L_\text{peak}}$. Thus the $L_\text{peak}$ is $6.6\times10^{44}$, $9.5\times10^{44}$ and $1.0\times10^{44}$~\lum\ for Si, SiC and graphite dust grains respectively. In the calculation we have taken the lower limit of the inner radius $R$ (1.2 pc, see \S\ref{modelfit}), so the obtained $L_\text{peak}$ can be considered as a lower limit.

Notably, all of the $L_\text{peak}$ values are significantly higher than the peak luminosity $2.7\times10^{43}$~\lum\ fitted by the optical-UV blackbody model~\citep{Hammerstein2023}. This discrepancy can serve as a compelling evidence for missing energy in the extreme-UV (EUV) band~\citep{Lu2018}, which is invisible directly but can be probed by dust echo emission.
Given a \mbh\ of $1.6\times10^{6}$\msun~\citep{Nicholl2020}, the corresponding lower limits of the Eddington ratios are 3.3, 4.7, and 0.49, respectively. We caution, however, that the derived $L_\text{peak}$ values are subject to considerable uncertainty, due to insufficient information on dust grain properties, such as grain size and the dust composition. The MIR spectroscopic observations with the James Webb Space Telescope (JWST) may refine our understanding of the grain properties, leading to a more accurate estimate of $L_{\text{peak}}$ and the missing energy.

\subsection{The link between IR bright TDEs and QPEs}
\label{IRTDE-QPE}

\begin{figure}
\centering
\begin{minipage}{0.5\textwidth}
\centering{\includegraphics[width=1\textwidth]{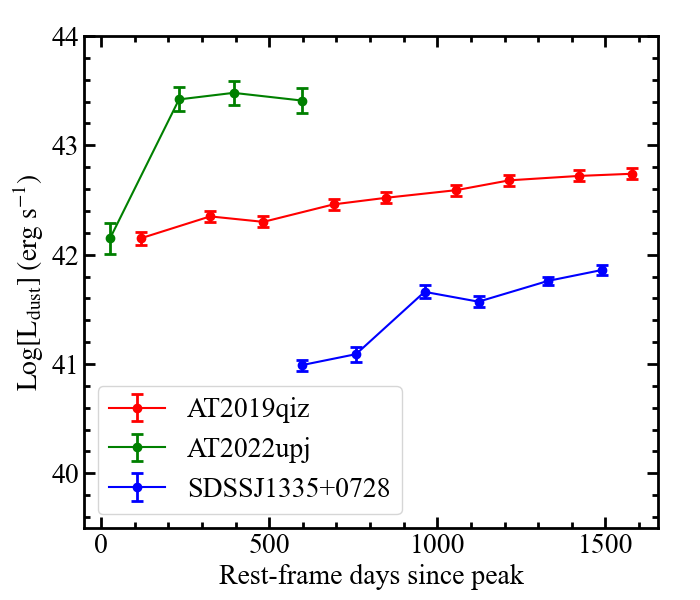}}
\end{minipage}
\caption{The dust luminosity evolution of QPE-associated TDE AT 2019qiz, AT2022upj and TDE candidate SDSSJ1335+0728. The dust luminosity is derived in the same way as \citet{Jiang2021}, assuming silicate grains with a size distribution of
$N(a)\propto a^{-3.5}$ and taking into account the absorption coefficient. The rest-frame days since peak are relative to the date of the optical flare peak (MJD $\sim58767.61$, $\sim59885$, and $\sim58991$, respectively).}
\label{luminosity_3}
\end{figure}

Optical TDEs rarely show bright IR echoes due to the lack of parsec-scale dust~\citep{Jiang2021}. However, AT 2019qiz, the first optical TDE with unambiguous detection of QPEs~\citep{Nicholl2024Natur}, happens to belong to the rare IR-bright subset, probably due to the presence of a torus remnant (see \S\ref{Large-scale-dust-torus}). More recently, QPEs have been reported in the other optical TDE AT2022upj~\citep{Chakraborty2025}. 
Again, AT2022upj is an IR-bright TDE ($L_{\rm dust}>10^{43}$~\lum, see Figure~\ref{luminosity_3}), and its circumnuclear dust has been inferred to be located at a distance of at least 0.4~pc~\citep{Newsome2024-upj}.
This intriguing coincidence strongly suggests that QPEs preferentially follow TDEs with prominent IR echoes. 
According to the unified model proposed by \cite{Jiang2025embers}, recently faded AGNs, TDEs and QPEs are all closely related. An earlier AGN phase has enhanced both the TDE rate \citep{Karas2007, Kennedy2016, Kaur2025} and the formation rate of low-eccentricity EMRIs~\citep{Pan2021prd,Zhou2024b}. Due to observational selection effects, TDEs are more likely to be identified in recently faded AGNs rather than in actively accreting ones. Consequently, QPEs are also preferentially found in these recently faded AGNs, where misaligned accretion disks formed from TDEs frequently feed the quasi-circular EMRIs.  In this scenario, a bright IR echo can naturally be addressed by the dusty torus left behind by a past AGN. 
In \S\ref{Large-scale-dust-torus}, we have calculated the dust covering factor, $f_c$, for AT 2019qiz from the modeled geometric parameters of the torus remnant.  Actually, the $f_c$ of an AGN torus has been widely measured from the ratio of IR to bolometric luminosity, yielding values of $\mathcal{O} (10^{-1})$~\citep{Maiolino2007,Mor2009,Roseboom2013}. Similarly, \citet{Jiang2021} used the ratio of the peak dust luminosity to the OUV bolometric luminosity of TDEs to estimate the $f_c$ and inferred very low values ($\lesssim 10^{-2}$) for inactive SMBHs. Using this method, a torus-like $f_c$ value of 0.48 was immediately obtained for AT2022upj, which is distinctive of normal optical TDEs but similar to AT 2019qiz. However, both AT 2019qiz and AT2022upj have torus inner radius much larger than those expected in the AGN phase, that is consistent with a fading torus scenario (see \S\ref{Large-scale-dust-torus}). 

We note another remarkable case of QPEs recently discovered in the galaxy SDSS J1335+0728~\citep{HernandezGarcia2025}, which is an inactive galaxy that exhibited an optical brightening (identified as ZTF19acnskyy) followed by persistent AGN-like variability since then~\citep{SanchezSaezet2024}. 
Although they suggested that QPEs in this galaxy are associated with a turn-on AGN, an exotic TDE scenario of ZTF19acnskyy cannot be convincingly ruled out given its distinct features consistent with TDEs, including a blue optical colour ($g-r\approx-0.5$), the soft X-ray emission ($kT_{\rm bb}\approx50-100$~eV) and the non-detection of broad emission in the optical spectra. As already noted in~\citet{SanchezSaezet2024}, the IR echo of ZTF19acnskyy shows an atypically long time delay (several years, see Figure~\ref{luminosity_3}), which is much larger than expected from the UV/optical emission reprocessing by a preexisting dusty torus around the SMBH. Nevertheless, the large dust structure is strikingly similar to that in AT 2019qiz, which may be a torus remnant. Further integral field spectrograph observations of the host of SDSSJ1335+0728 will be able to test whether or not there is an EELR supporting a recently faded AGN.  We emphasize that the relatively low IR luminosity observed in SDSSJ1335+0728 is due to a relatively low OUV luminosity ($\sim10^{43}$~\lum, \citealt{SanchezSaezet2024}) and a still rising trend.

The universal detection of bright IR echoes in QPE-associated TDEs or TDE candidates not only provides a new clue for understanding the formation of QPEs, but also offers a new approach for searching for new QPEs. At present, long-term and high-cadence X-ray observations of all TDEs and associated nuclear transients remain unrealistic due to limited X-ray observational resources. In contrast, the number of IR-bright TDEs is much smaller and the IR echoes can be measured by MIR or NIR observations, which are feasible thanks to a much larger number of facilities in these bands. 

\section{Conclusion}

We presented a comprehensive analysis of the strong MIR echo found in AT 2019qiz, the first optical TDE with clear QPE detection. Following the discovery of AT 2019qiz, a total of 9 brightened epochs have been detected in the half-yearly cadence IR light curves based on WISE and NEOWISE archive data up to 2024 February, the last visit before the retirement of the WISE satellite. The IR light curves exhibit an early bump, followed by a steady rise until the penultimate epoch, after which it appears to enter a plateau phase. The inferred dust temperature decreases until the fourth epoch and then remains approximately constant for the following five epochs. Throughout the entire period of IR detection, the dust temperature remains below the sublimation threshold.

The intriguing behaviour of the IR light curves of AT 2019qiz is reminiscent of the long-lasting IR flares found in IRAS~F01004-2237~\citep{Dou2017} and PS16dtm~\citep{Jiang2025PS16dtm}, which can be well modeled by the echo of a torus with a fixed but large inner radius. In this scenario, the IR emission with an almost constant temperature is due to the dominant emission of the dust at the inner surface, while it is distributed on a yearly time scale due to the time delay effect of the dust at different positions. Our fit shows that the emitted dust is located at a distance of about 1~pc from the SMBH, with little dependence on the composition and size of the dust grains. A torus with such a large radius is unusual for an SMBH with a mass of $\sim10^6$~\msun\, since it is far beyond the inner radius determined by dust sublimation, even assuming Eddington accretion. Combined with the recent discovery of an EELR in the host of AT 2019qiz~\citep{Xiong2025}, we prefer to attribute the large radius to a receding torus from a past AGN activity.  Interestingly, such a large torus-like dust structure is also tentatively uncoverved by its bright and long IR echoes in two other QPE sources, following an optical TDE AT2022upj and an exotic TDE candidate in SDSSJ1335+0728, respectively. Thus the torus remnant may also serve as another independent evidence for a recently faded AGN, which could be an essential process to trigger both the TDE and the QPEs that subsequently occurred~\citep{Jiang2025embers}. In this scenario, IR-bright TDEs could be the most promising sources associated with QPEs, and opens a new way to detect QPEs, i.e. by monitoring TDEs with obvious dust echoes at high X-ray cadence. It should be noted that although the assumption in \citet{Pasham2025} is not applicable to AT 2019qiz, it remains possible that the IR emission from dust heated by QPEs in other TDEs is significant and represents another factor linking QPEs and IR-bright TDEs.

The other implication learned from our dust echo model is the high bolometric luminosity required to heat the dust, with a peak value much higher than that fitted from the OUV blackbody, i.e. $6.6\times10^{44}$~\lum\ in the case of 0.1~$\mu$m silicate dust grains. This suggests that AT 2019qiz is probably undergoing super-Eddington accretion during its peak, while most of its energy has been radiated in the form of unobserved EUV photons, providing a solution to the so-called missing energy puzzle. Due to the lack of IR observations during the decline phase of the light curve, our model currently only provides a lower limit on the inner radius of the torus. Although the NEOWISE mission has now ended, observations with its successor, the NEO Surveyor \citep{Mainzer2014}, as well as the Nancy Grace Roman Space Telescope \citep{Spergel2015}, may detect IR echoes from AT 2019qiz in the near future and help to further constrain the geometry and evolution of the dust torus. In addition, JWST spectroscopic observations could provide critical insights into the dust composition and grain properties, ultimately refining our estimates of the total radiative output of the TDE.

\section{ACKNOWLEDGMENTS}
We thank the referee for very positive and constructive comments, which have improved the manuscript significantly. We also thank Wenbin Lu, Luming Sun and Zhen Pan for their helpful discussions.
This work is supported by the National Key Research and Development Program of China (2023YFA1608100), the Strategic Priority Research Program of the Chinese Academy of Sciences
(XDB0550200), the National Natural Science Foundation
of China (grants 12192221,12393814) and the China Manned Space Project.
M.W. gratefully acknowledge the support of the National Undergraduate Training Program for Innovation and Entrepreneurship. This research makes use of data products from the Wide-field Infrared Survey Explorer, which is a joint project of the University of California, Los Angeles, and the Jet Propulsion Laboratory/California Institute of Technology, funded by the National Aeronautics and Space Administration. This research also makes use of data products from NEOWISE-R, which is a project of the Jet Propulsion Laboratory/California Institute of Technology, funded by the Planetary Science Division of the National Aeronautics and Space Administration. This research has made use of the NASA/IPAC Infrared Science Archive, which is operated by the California Institute of Technology, under contract with the National Aeronautics and Space Administration.

\appendix
\counterwithin{figure}{section}

\section{Appendix Figures} \label{appendix:A}
\begin{figure*}
\centering
\begin{minipage}{1.0\textwidth}
\centering{\includegraphics[width=0.9\textwidth]{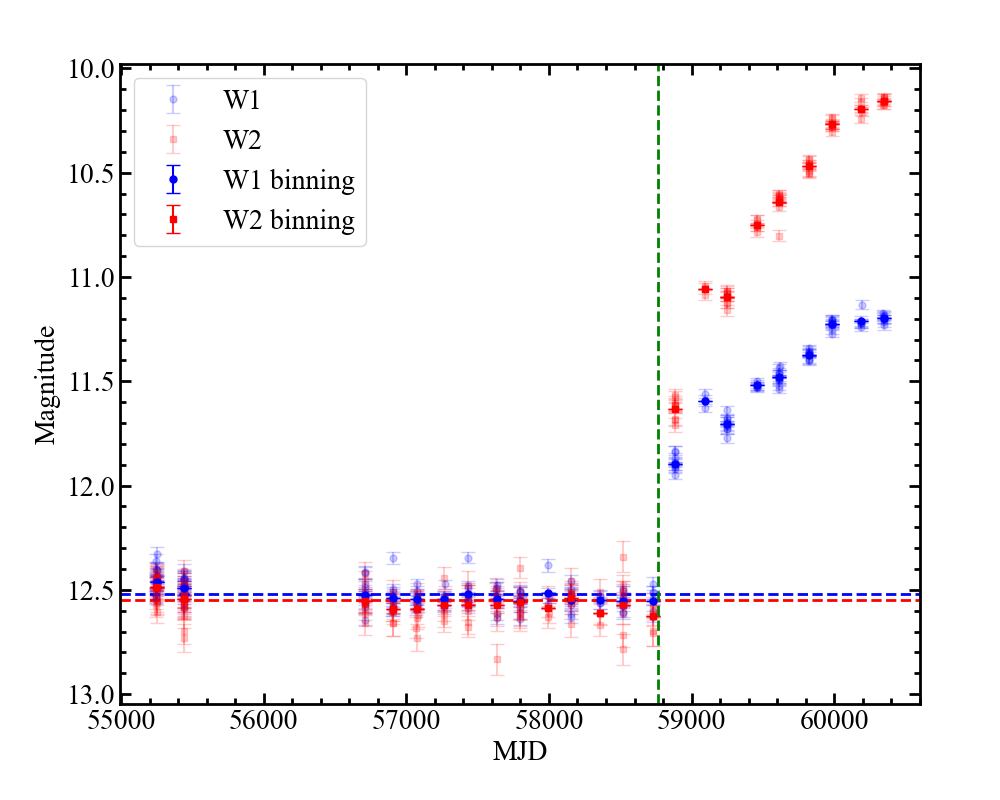}}
\end{minipage}
\caption{The WISE light curves of AT 2019qiz. Blue dots: W1 ($3.4\mu$m); red squares: W2 ($4.6\mu$m). The raw single exposures are plotted in light blue and red, while the binned data are plotted in dark blue and red.
The red and blue dashed lines denote the average magnitudes at the quiescent phase. The green dashed line marks the time of optical peak.}
\label{WISE}
\end{figure*}

\begin{figure*}
\centering
\begin{minipage}{1.0\textwidth}
\centering{\includegraphics[width=0.9\textwidth]{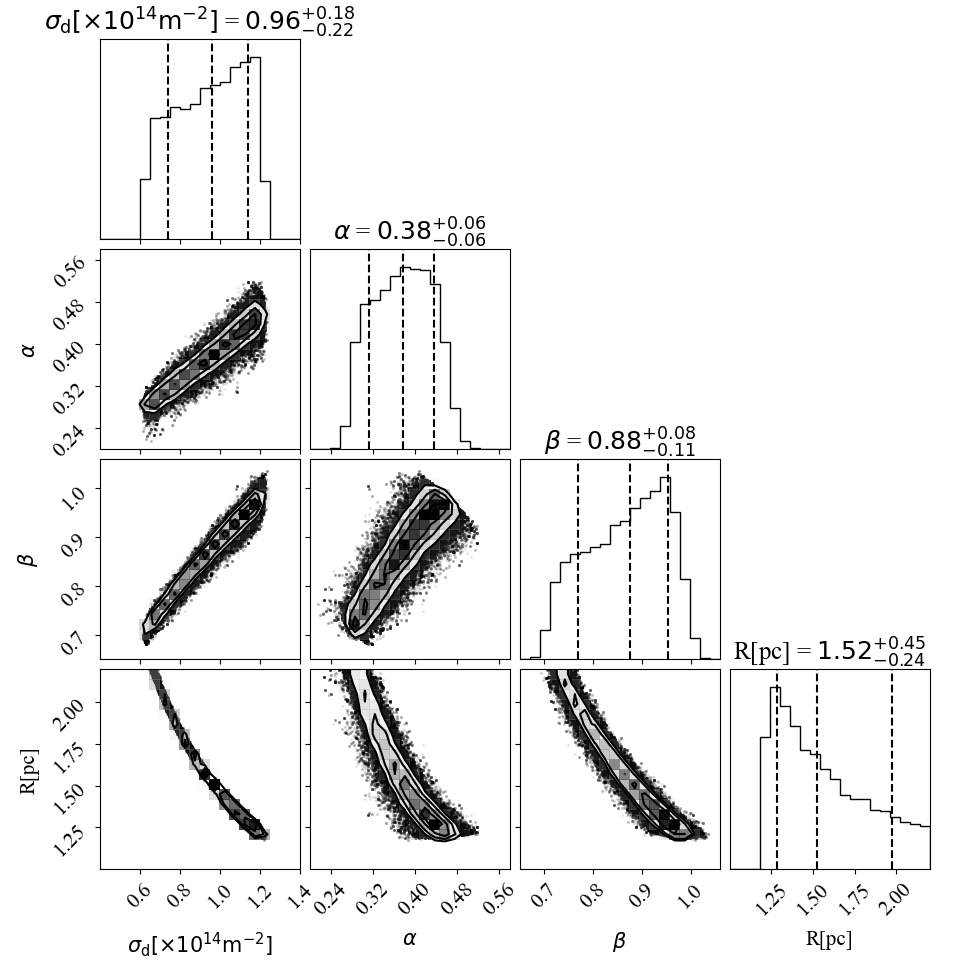}}
\end{minipage}
\caption{Similar to Figure~\ref{mcmcsi}, but for SiC dust grains, showing the distributions of the dust echo parameters.}
\label{mcmcsic}
\end{figure*}

\begin{figure*}
\centering
\begin{minipage}{1.0\textwidth}
\centering{\includegraphics[width=0.9\textwidth]{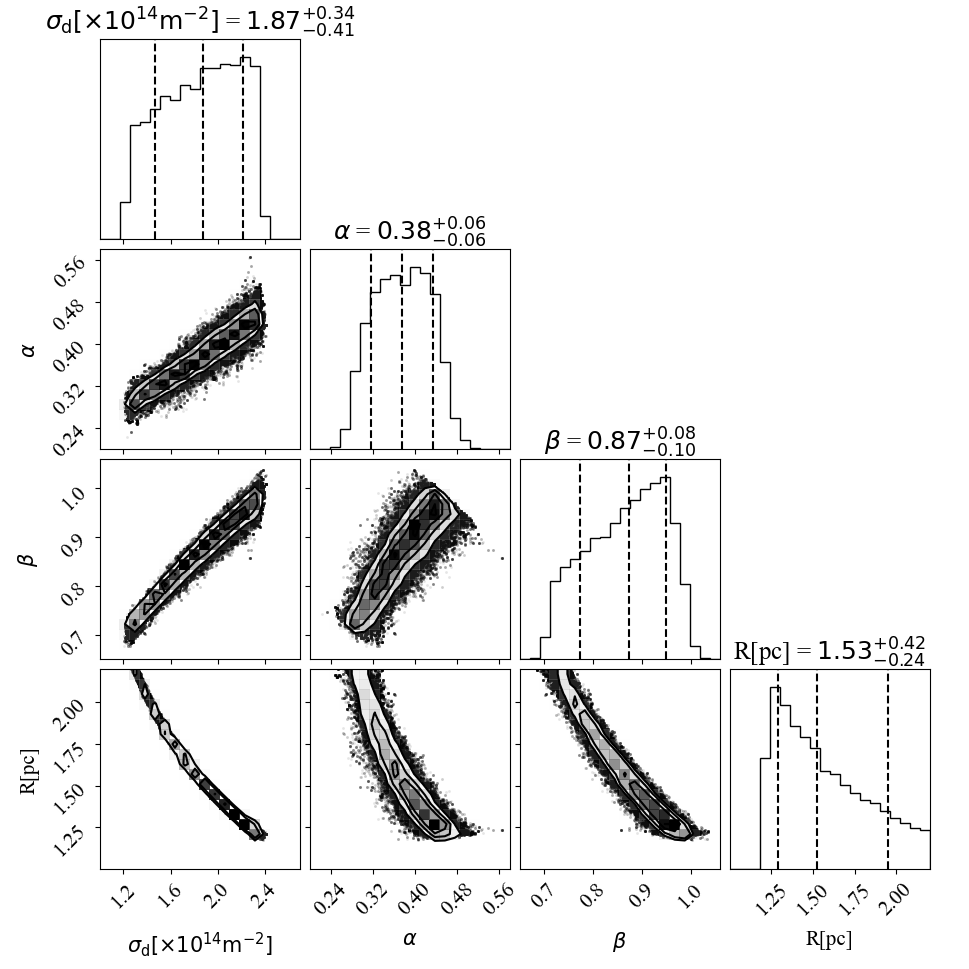}}
\end{minipage}
\caption{Similar to Figure~\ref{mcmcsi}, but for graphite dust grains, showing the distributions of the dust echo parameters.}
\label{mcmcgra}
\end{figure*}

\begin{figure*}
\centering
\begin{minipage}{1.0\textwidth}
\centering{\includegraphics[width=0.9\textwidth]{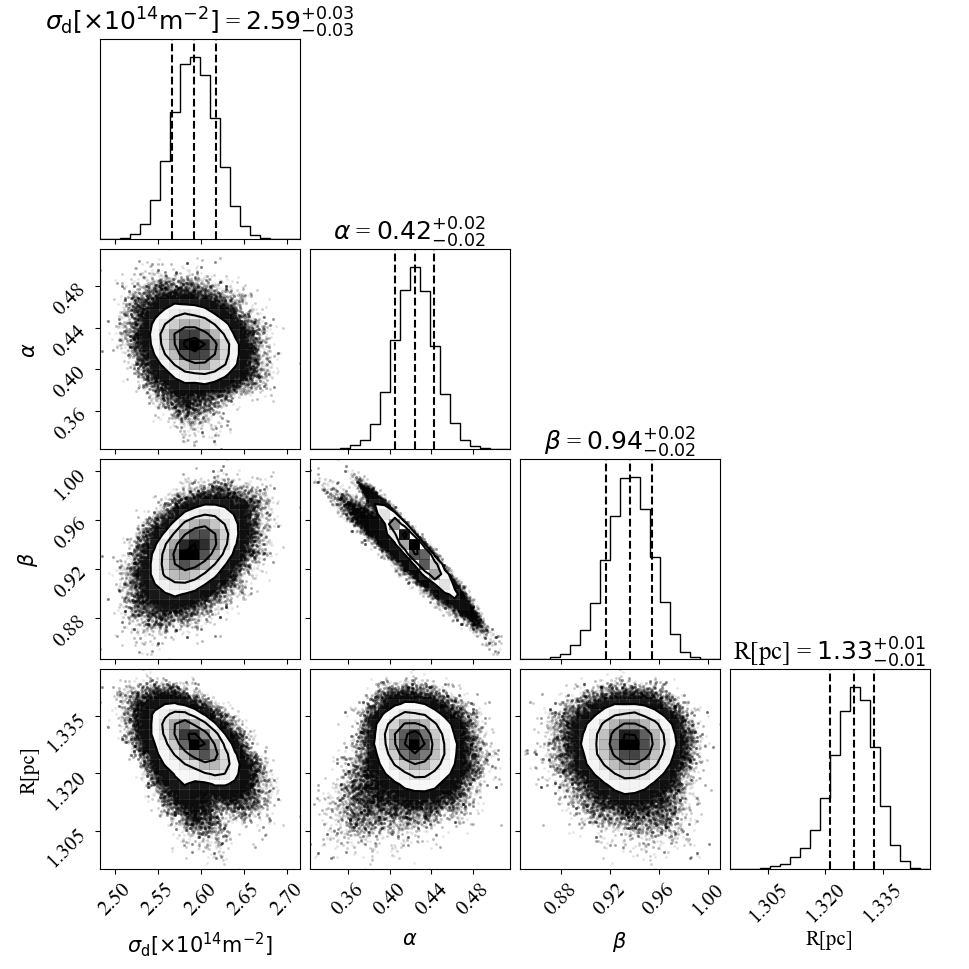}}
\end{minipage}
\caption{This figure is similar to Figure~\ref{mcmcsi}, but with a set of IR data artificially added at MJD=60700 for the fitting process, showing the distributions of the dust echo parameters.}
\label{add}
\end{figure*}

\begin{figure*}
\centering
\begin{minipage}{1.0\textwidth}
\centering{\includegraphics[width=0.9\textwidth]{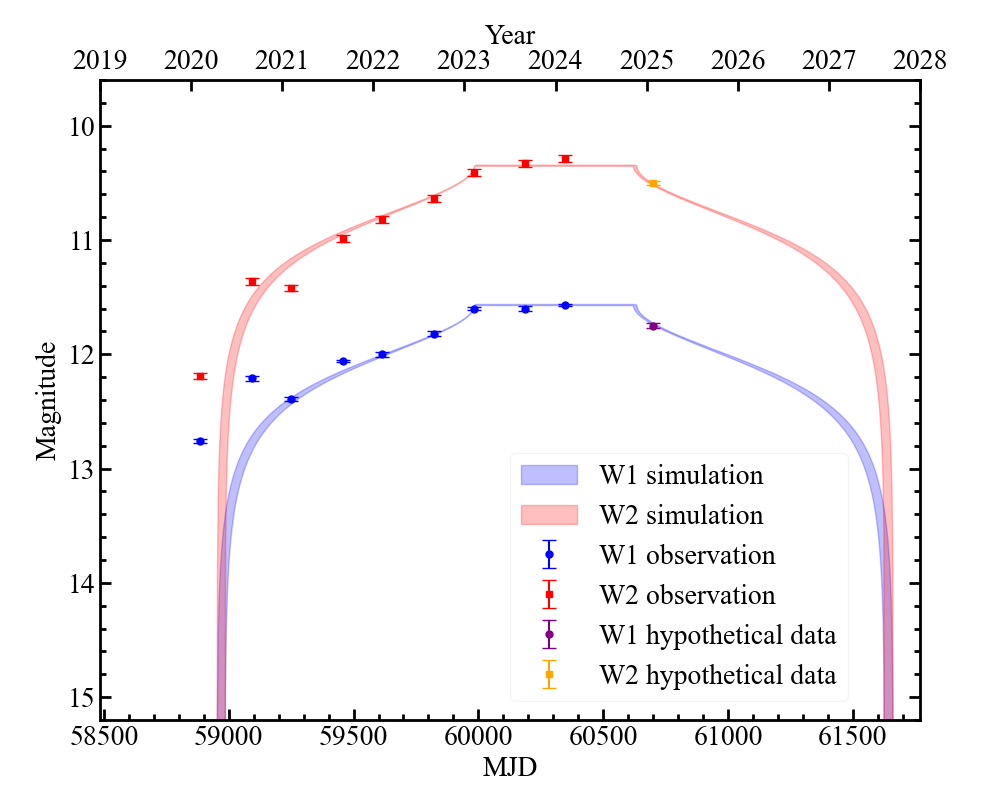}}
\end{minipage}
\caption{This figure is similar to Figure~\ref{lightcurve}, but with a set of data artificially added at MJD=60700 for the fitting process, showing the simulated IR light curves. The purple and orange dots denote the hypothetical W1 and W2 data in the decline phase, respectively.}
\label{lc_add}
\end{figure*}

In this section, we present five supplementary figures referenced in the main text that facilitate comprehensive understanding without disrupting the narrative flow.
The WISE light curves of AT 2019qiz are shown in Figure~\ref{WISE}. The posterior samples distributions of SiC, Gra are shown in Figure~\ref{mcmcsic} and~\ref{mcmcgra}, respectively. The posterior sample distributions and the simulated IR light curves obtained by incorporating a set of artificially added data are shown in Figure~\ref{add} and Figure~\ref{lc_add}, respectively.

\clearpage

\bibliographystyle{aasjournal}
\bibliography{reference}

\begin{thebibliography}{}
\expandafter\ifx\csname natexlab\endcsname\relax\def\natexlab#1{#1}\fi
\providecommand{\url}[1]{\href{#1}{#1}}
\providecommand{\dodoi}[1]{doi:~\href{http://doi.org/#1}{\nolinkurl{#1}}}
\providecommand{\doeprint}[1]{\href{http://ascl.net/#1}{\nolinkurl{http://ascl.net/#1}}}
\providecommand{\doarXiv}[1]{\href{https://arxiv.org/abs/#1}{\nolinkurl{https://arxiv.org/abs/#1}}}

\bibitem[{{Arcodia} {et~al.}(2021){Arcodia}, {Merloni}, {Nandra}, {Buchner}, {Salvato}, {Pasham}, {Remillard}, {Comparat}, {Lamer}, {Ponti}, {Malyali}, {Wolf}, {Arzoumanian}, {Bogensberger}, {Buckley}, {Gendreau}, {Gromadzki}, {Kara}, {Krumpe}, {Markwardt}, {Ramos-Ceja}, {Rau}, {Schramm}, \& {Schwope}}]{Arcodia2021}
{Arcodia}, R., {Merloni}, A., {Nandra}, K., {et~al.} 2021, \nat, 592, 704, \dodoi{10.1038/s41586-021-03394-6}

\bibitem[{{Balokovi{\'c}} {et~al.}(2018){Balokovi{\'c}}, {Brightman}, {Harrison}, {Comastri}, {Ricci}, {Buchner}, {Gandhi}, {Farrah}, \& {Stern}}]{Balokovic2018}
{Balokovi{\'c}}, M., {Brightman}, M., {Harrison}, F.~A., {et~al.} 2018, \apj, 854, 42, \dodoi{10.3847/1538-4357/aaa7eb}

\bibitem[{{Beckert} \& {Duschl}(2004)}]{Beckert2004}
{Beckert}, T., \& {Duschl}, W.~J. 2004, \aap, 426, 445, \dodoi{10.1051/0004-6361:20040336}

\bibitem[{{Chakraborty} {et~al.}(2025){Chakraborty}, {Kara}, {Arcodia}, {Buchner}, {Giustini}, {Hern{\'a}ndez-Garc{\'\i}a}, {Linial}, {Masterson}, {Miniutti}, {Mummery}, {Panagiotou}, {Quintin}, \& {S{\'a}nchez-S{\'a}ez}}]{Chakraborty2025}
{Chakraborty}, J., {Kara}, E., {Arcodia}, R., {et~al.} 2025, \apjl, 983, L39, \dodoi{10.3847/2041-8213/adc2f8}

\bibitem[{{Chan} \& {Krolik}(2016)}]{Chan2016}
{Chan}, C.-H., \& {Krolik}, J.~H. 2016, \apj, 825, 67, \dodoi{10.3847/0004-637X/825/1/67}

\bibitem[{{Dou} {et~al.}(2017){Dou}, {Wang}, {Yan}, {Jiang}, {Yang}, {Cutri}, {Mainzer}, \& {Peng}}]{Dou2017}
{Dou}, L., {Wang}, T., {Yan}, L., {et~al.} 2017, \apjl, 841, L8, \dodoi{10.3847/2041-8213/aa7130}

\bibitem[{{Elitzur} \& {Ho}(2009)}]{Elitzur2009}
{Elitzur}, M., \& {Ho}, L.~C. 2009, \apjl, 701, L91, \dodoi{10.1088/0004-637X/701/2/L91}

\bibitem[{{Elitzur} \& {Shlosman}(2006)}]{Elitzur2006}
{Elitzur}, M., \& {Shlosman}, I. 2006, \apjl, 648, L101, \dodoi{10.1086/508158}

\bibitem[{{Emmering} {et~al.}(1992){Emmering}, {Blandford}, \& {Shlosman}}]{Emmering1992}
{Emmering}, R.~T., {Blandford}, R.~D., \& {Shlosman}, I. 1992, \apj, 385, 460, \dodoi{10.1086/170955}

\bibitem[{{Foreman-Mackey} {et~al.}(2013){Foreman-Mackey}, {Hogg}, {Lang}, \& {Goodman}}]{2013PASP..125..306F}
{Foreman-Mackey}, D., {Hogg}, D.~W., {Lang}, D., \& {Goodman}, J. 2013, \pasp, 125, 306, \dodoi{10.1086/670067}

\bibitem[{{Gezari}(2021)}]{Gezari2021}
{Gezari}, S. 2021, \araa, 59, 21, \dodoi{10.1146/annurev-astro-111720-030029}

\bibitem[{{Gilbert} {et~al.}(2024){Gilbert}, {Ruan}, {Eracleous}, {Haggard}, \& {Runnoe}}]{Gilbert2024}
{Gilbert}, O., {Ruan}, J.~J., {Eracleous}, M., {Haggard}, D., \& {Runnoe}, J.~C. 2024, arXiv e-prints, arXiv:2409.10486, \dodoi{10.48550/arXiv.2409.10486}

\bibitem[{{Hamann} {et~al.}(1993){Hamann}, {Korista}, \& {Morris}}]{Hamann1993}
{Hamann}, F., {Korista}, K.~T., \& {Morris}, S.~L. 1993, \apj, 415, 541, \dodoi{10.1086/173185}

\bibitem[{{Hammerstein} {et~al.}(2023){Hammerstein}, {van Velzen}, {Gezari}, {Cenko}, {Yao}, {Ward}, {Frederick}, {Villanueva}, {Somalwar}, {Graham}, {Kulkarni}, {Stern}, {Andreoni}, {Bellm}, {Dekany}, {Dhawan}, {Drake}, {Fremling}, {Gatkine}, {Groom}, {Ho}, {Kasliwal}, {Karambelkar}, {Kool}, {Masci}, {Medford}, {Perley}, {Purdum}, {van Roestel}, {Sharma}, {Sollerman}, {Taggart}, \& {Yan}}]{Hammerstein2023}
{Hammerstein}, E., {van Velzen}, S., {Gezari}, S., {et~al.} 2023, \apj, 942, 9, \dodoi{10.3847/1538-4357/aca283}

\bibitem[{{Hern{\'a}ndez-Garc{\'\i}a} {et~al.}(2025){Hern{\'a}ndez-Garc{\'\i}a}, {Chakraborty}, {S{\'a}nchez-S{\'a}ez}, {Ricci}, {Cuadra}, {McKernan}, {Ford}, {Ar{\'e}valo}, {Rau}, {Arcodia}, {Kara}, {Liu}, {Merloni}, {Bruni}, {Goodwin}, {Arzoumanian}, {Assef}, {Baldini}, {Bayo}, {Bauer}, {Bernal}, {Brightman}, {Calistro Rivera}, {Gendreau}, {Homan}, {Krumpe}, {Lira}, {Mart{\'\i}nez-Aldama}, {Salvato}, \& {Sotomayor}}]{HernandezGarcia2025}
{Hern{\'a}ndez-Garc{\'\i}a}, L., {Chakraborty}, J., {S{\'a}nchez-S{\'a}ez}, P., {et~al.} 2025, Nature Astronomy, 9, 895, \dodoi{10.1038/s41550-025-02523-9}

\bibitem[{{Hopkins} {et~al.}(2016){Hopkins}, {Torrey}, {Faucher-Gigu{\`e}re}, {Quataert}, \& {Murray}}]{Hopkins2016}
{Hopkins}, P.~F., {Torrey}, P., {Faucher-Gigu{\`e}re}, C.-A., {Quataert}, E., \& {Murray}, N. 2016, \mnras, 458, 816, \dodoi{10.1093/mnras/stw289}

\bibitem[{{Hsieh} {et~al.}(2021){Hsieh}, {Koch}, {Kim}, {Mart{\'\i}n}, {Yen}, {Carpenter}, {Harada}, {Turner}, {Ho}, {Tang}, \& {Beck}}]{Hsieh2021}
{Hsieh}, P.-Y., {Koch}, P.~M., {Kim}, W.-T., {et~al.} 2021, \apj, 913, 94, \dodoi{10.3847/1538-4357/abf4cd}

\bibitem[{{Jiang} {et~al.}(2016){Jiang}, {Dou}, {Wang}, {Yang}, {Lyu}, \& {Zhou}}]{Jiang2016}
{Jiang}, N., {Dou}, L., {Wang}, T., {et~al.} 2016, \apjl, 828, L14, \dodoi{10.3847/2041-8205/828/1/L14}

\bibitem[{{Jiang} {et~al.}(2025){Jiang}, {Luo}, {Zhu}, \& {Cutri}}]{Jiang2025PS16dtm}
{Jiang}, N., {Luo}, D., {Zhu}, J., \& {Cutri}, R.~M. 2025, \apjl, 980, L17, \dodoi{10.3847/2041-8213/adaeb9}

\bibitem[{{Jiang} \& {Pan}(2025)}]{Jiang2025embers}
{Jiang}, N., \& {Pan}, Z. 2025, \apjl, 983, L18, \dodoi{10.3847/2041-8213/adc456}

\bibitem[{{Jiang} {et~al.}(2021{\natexlab{a}}){Jiang}, {Wang}, {Hu}, {Sun}, {Dou}, \& {Xiao}}]{Jiang2021}
{Jiang}, N., {Wang}, T., {Hu}, X., {et~al.} 2021{\natexlab{a}}, \apj, 911, 31, \dodoi{10.3847/1538-4357/abe772}

\bibitem[{{Jiang} {et~al.}(2019){Jiang}, {Wang}, {Mou}, {Liu}, {Dou}, {Sheng}, \& {Wang}}]{Jiang2019}
{Jiang}, N., {Wang}, T., {Mou}, G., {et~al.} 2019, \apj, 871, 15, \dodoi{10.3847/1538-4357/aaf6b2}

\bibitem[{{Jiang} {et~al.}(2021{\natexlab{b}}){Jiang}, {Wang}, {Dou}, {Shu}, {Hu}, {Liu}, {Wang}, {Yan}, {Sheng}, {Yang}, {Sun}, \& {Zhou}}]{Jiang2021a}
{Jiang}, N., {Wang}, T., {Dou}, L., {et~al.} 2021{\natexlab{b}}, \apjs, 252, 32, \dodoi{10.3847/1538-4365/abd1dc}

\bibitem[{{Karas} \& {{\v{S}}ubr}(2007)}]{Karas2007}
{Karas}, V., \& {{\v{S}}ubr}, L. 2007, \aap, 470, 11, \dodoi{10.1051/0004-6361:20066068}

\bibitem[{{Kaur} \& {Stone}(2025)}]{Kaur2025}
{Kaur}, K., \& {Stone}, N.~C. 2025, \apj, 979, 172, \dodoi{10.3847/1538-4357/ad9b86}

\bibitem[{{Kennedy} {et~al.}(2016){Kennedy}, {Meiron}, {Shukirgaliyev}, {Panamarev}, {Berczik}, {Just}, \& {Spurzem}}]{Kennedy2016}
{Kennedy}, G.~F., {Meiron}, Y., {Shukirgaliyev}, B., {et~al.} 2016, \mnras, 460, 240, \dodoi{10.1093/mnras/stw908}

\bibitem[{{Konigl} \& {Kartje}(1994)}]{Konigl1994}
{Konigl}, A., \& {Kartje}, J.~F. 1994, \apj, 434, 446, \dodoi{10.1086/174746}

\bibitem[{{Koshida} {et~al.}(2014){Koshida}, {Minezaki}, {Yoshii}, {Kobayashi}, {Sakata}, {Sugawara}, {Enya}, {Suganuma}, {Tomita}, {Aoki}, \& {Peterson}}]{Koshida2014}
{Koshida}, S., {Minezaki}, T., {Yoshii}, Y., {et~al.} 2014, \apj, 788, 159, \dodoi{10.1088/0004-637X/788/2/159}

\bibitem[{{Krolik}(2007)}]{Krolik2007}
{Krolik}, J.~H. 2007, \apj, 661, 52, \dodoi{10.1086/515432}

\bibitem[{{Krolik} \& {Begelman}(1988)}]{Krolik1988}
{Krolik}, J.~H., \& {Begelman}, M.~C. 1988, \apj, 329, 702, \dodoi{10.1086/166414}

\bibitem[{{Laor} \& {Draine}(1993)}]{Laor1993}
{Laor}, A., \& {Draine}, B.~T. 1993, \apj, 402, 441, \dodoi{10.1086/172149}

\bibitem[{{Latvakoski} {et~al.}(1999){Latvakoski}, {Stacey}, {Gull}, \& {Hayward}}]{Latvakoski1999}
{Latvakoski}, H.~M., {Stacey}, G.~J., {Gull}, G.~E., \& {Hayward}, T.~L. 1999, \apj, 511, 761, \dodoi{10.1086/306689}

\bibitem[{{Lau} {et~al.}(2013){Lau}, {Herter}, {Morris}, {Becklin}, \& {Adams}}]{Lau2013}
{Lau}, R.~M., {Herter}, T.~L., {Morris}, M.~R., {Becklin}, E.~E., \& {Adams}, J.~D. 2013, \apj, 775, 37, \dodoi{10.1088/0004-637X/775/1/37}

\bibitem[{{Lin} {et~al.}(2022){Lin}, {Jiang}, {Kong}, {Huang}, {Lin}, {Zhu}, \& {Wang}}]{Lin2022}
{Lin}, Z., {Jiang}, N., {Kong}, X., {et~al.} 2022, \apjl, 939, L33, \dodoi{10.3847/2041-8213/ac9c63}

\bibitem[{{Linial} \& {Metzger}(2023)}]{Linial2023}
{Linial}, I., \& {Metzger}, B.~D. 2023, \apj, 957, 34, \dodoi{10.3847/1538-4357/acf65b}

\bibitem[{{Lu} \& {Kumar}(2018)}]{Lu2018}
{Lu}, W., \& {Kumar}, P. 2018, \apj, 865, 128, \dodoi{10.3847/1538-4357/aad54a}

\bibitem[{{Lu} {et~al.}(2016){Lu}, {Kumar}, \& {Evans}}]{Lu2016}
{Lu}, W., {Kumar}, P., \& {Evans}, N.~J. 2016, \mnras, 458, 575, \dodoi{10.1093/mnras/stw307}

\bibitem[{{Lyu} {et~al.}(2019){Lyu}, {Rieke}, \& {Smith}}]{Lyu2019}
{Lyu}, J., {Rieke}, G.~H., \& {Smith}, P.~S. 2019, \apj, 886, 33, \dodoi{10.3847/1538-4357/ab481d}

\bibitem[{{Mainzer} {et~al.}(2014){Mainzer}, {Bauer}, {Cutri}, {Grav}, {Masiero}, {Beck}, {Clarkson}, {Conrow}, {Dailey}, {Eisenhardt}, {Fabinsky}, {Fajardo-Acosta}, {Fowler}, {Gelino}, {Grillmair}, {Heinrichsen}, {Kendall}, {Kirkpatrick}, {Liu}, {Masci}, {McCallon}, {Nugent}, {Papin}, {Rice}, {Royer}, {Ryan}, {Sevilla}, {Sonnett}, {Stevenson}, {Thompson}, {Wheelock}, {Wiemer}, {Wittman}, {Wright}, \& {Yan}}]{Mainzer2014}
{Mainzer}, A., {Bauer}, J., {Cutri}, R.~M., {et~al.} 2014, \apj, 792, 30, \dodoi{10.1088/0004-637X/792/1/30}

\bibitem[{{Maiolino} {et~al.}(2007){Maiolino}, {Shemmer}, {Imanishi}, {Netzer}, {Oliva}, {Lutz}, \& {Sturm}}]{Maiolino2007}
{Maiolino}, R., {Shemmer}, O., {Imanishi}, M., {et~al.} 2007, \aap, 468, 979, \dodoi{10.1051/0004-6361:20077252}

\bibitem[{{Masci} {et~al.}(2019){Masci}, {Laher}, {Rusholme}, {Shupe}, {Groom}, {Surace}, {Jackson}, {Monkewitz}, {Beck}, {Flynn}, {Terek}, {Landry}, {Hacopians}, {Desai}, {Howell}, {Brooke}, {Imel}, {Wachter}, {Ye}, {Lin}, {Cenko}, {Cunningham}, {Rebbapragada}, {Bue}, {Miller}, {Mahabal}, {Bellm}, {Patterson}, {Juri{\'c}}, {Golkhou}, {Ofek}, {Walters}, {Graham}, {Kasliwal}, {Dekany}, {Kupfer}, {Burdge}, {Cannella}, {Barlow}, {Van Sistine}, {Giomi}, {Fremling}, {Blagorodnova}, {Levitan}, {Riddle}, {Smith}, {Helou}, {Prince}, \& {Kulkarni}}]{Masci2019}
{Masci}, F.~J., {Laher}, R.~R., {Rusholme}, B., {et~al.} 2019, \pasp, 131, 018003, \dodoi{10.1088/1538-3873/aae8ac}

\bibitem[{{Masterson} {et~al.}(2024){Masterson}, {De}, {Panagiotou}, {Kara}, {Arcavi}, {Eilers}, {Frostig}, {Gezari}, {Grotova}, {Liu}, {Malyali}, {Meisner}, {Merloni}, {Newsome}, {Rau}, {Simcoe}, \& {van Velzen}}]{Masterson2024}
{Masterson}, M., {De}, K., {Panagiotou}, C., {et~al.} 2024, \apj, 961, 211, \dodoi{10.3847/1538-4357/ad18bb}

\bibitem[{{Masterson} {et~al.}(2025){Masterson}, {De}, {Panagiotou}, {Kara}, {Eilers}, {Guolo}, {Lu}, {Rest}, {Ricci}, \& {van Velzen}}]{Masterson2025}
---. 2025, arXiv e-prints, arXiv:2503.08647, \dodoi{10.48550/arXiv.2503.08647}

\bibitem[{{Miniutti} {et~al.}(2019){Miniutti}, {Saxton}, {Giustini}, {Alexander}, {Fender}, {Heywood}, {Monageng}, {Coriat}, {Tzioumis}, {Read}, {Knigge}, {Gandhi}, {Pretorius}, \& {Ag{\'\i}s-Gonz{\'a}lez}}]{Miniutti2019GSN069}
{Miniutti}, G., {Saxton}, R.~D., {Giustini}, M., {et~al.} 2019, \nat, 573, 381, \dodoi{10.1038/s41586-019-1556-x}

\bibitem[{{Mor} {et~al.}(2009){Mor}, {Netzer}, \& {Elitzur}}]{Mor2009}
{Mor}, R., {Netzer}, H., \& {Elitzur}, M. 2009, \apj, 705, 298, \dodoi{10.1088/0004-637X/705/1/298}

\bibitem[{{Necker} {et~al.}(2025){Necker}, {Graikou}, {Kowalski}, {Franckowiak}, {Nordin}, {Pernice}, {van Velzen}, \& {Veres}}]{Necker2025}
{Necker}, J., {Graikou}, E., {Kowalski}, M., {et~al.} 2025, \aap, 695, A228, \dodoi{10.1051/0004-6361/202451340}

\bibitem[{{Newsome} {et~al.}(2024){Newsome}, {Arcavi}, {Howell}, {McCully}, {Terreran}, {Hosseinzadeh}, {Bostroem}, {Dgany}, {Farah}, {Faris}, {Padilla-Gonzalez}, {Pellegrino}, \& {Andrews}}]{Newsome2024-upj}
{Newsome}, M., {Arcavi}, I., {Howell}, D.~A., {et~al.} 2024, \apj, 977, 258, \dodoi{10.3847/1538-4357/ad8a69}

\bibitem[{{Nicholl} {et~al.}(2020){Nicholl}, {Wevers}, {Oates}, {Alexander}, {Leloudas}, {Onori}, {Jerkstrand}, {Gomez}, {Campana}, {Arcavi}, {Charalampopoulos}, {Gromadzki}, {Ihanec}, {Jonker}, {Lawrence}, {Mandel}, {Schulze}, {Short}, {Burke}, {McCully}, {Hiramatsu}, {Howell}, {Pellegrino}, {Abbot}, {Anderson}, {Berger}, {Blanchard}, {Cannizzaro}, {Chen}, {Dennefeld}, {Galbany}, {Gonz{\'a}lez-Gait{\'a}n}, {Hosseinzadeh}, {Inserra}, {Irani}, {Kuin}, {M{\"u}ller-Bravo}, {Pineda}, {Ross}, {Roy}, {Smartt}, {Smith}, {Tucker}, {Wyrzykowski}, \& {Young}}]{Nicholl2020}
{Nicholl}, M., {Wevers}, T., {Oates}, S.~R., {et~al.} 2020, \mnras, 499, 482, \dodoi{10.1093/mnras/staa2824}

\bibitem[{{Nicholl} {et~al.}(2024){Nicholl}, {Pasham}, {Mummery}, {Guolo}, {Gendreau}, {Dewangan}, {Ferrara}, {Remillard}, {Bonnerot}, {Chakraborty}, {Hajela}, {Dhillon}, {Gillan}, {Greenwood}, {Huber}, {Janiuk}, {Salvesen}, {van Velzen}, {Aamer}, {Alexander}, {Angus}, {Arzoumanian}, {Auchettl}, {Berger}, {de Boer}, {Cendes}, {Chambers}, {Chen}, {Chornock}, {Fulton}, {Gao}, {Gillanders}, {Gomez}, {Gompertz}, {Fabian}, {Herman}, {Ingram}, {Kara}, {Laskar}, {Lawrence}, {Lin}, {Lowe}, {Magnier}, {Margutti}, {McGee}, {Minguez}, {Moore}, {Nathan}, {Oates}, {Patra}, {Ramsden}, {Ravi}, {Ridley}, {Sheng}, {Smartt}, {Smith}, {Srivastav}, {Stein}, {Stevance}, {Turner}, {Wainscoat}, {Weston}, {Wevers}, \& {Young}}]{Nicholl2024Natur}
{Nicholl}, M., {Pasham}, D.~R., {Mummery}, A., {et~al.} 2024, \nat, 634, 804, \dodoi{10.1038/s41586-024-08023-6}

\bibitem[{{Pan} \& {Yang}(2021)}]{Pan2021prd}
{Pan}, Z., \& {Yang}, H. 2021, \prd, 103, 103018, \dodoi{10.1103/PhysRevD.103.103018}

\bibitem[{{Pasham} {et~al.}(2025){Pasham}, {Coughlin}, {van Velzen}, \& {Hinkle}}]{Pasham2025}
{Pasham}, D.~R., {Coughlin}, E., {van Velzen}, S., \& {Hinkle}, J. 2025, arXiv e-prints, arXiv:2502.12078, \dodoi{10.48550/arXiv.2502.12078}

\bibitem[{{Petrushevska} {et~al.}(2023){Petrushevska}, {Leloudas}, {Ili{\'c}}, {Bronikowski}, {Charalampopoulos}, {Jaisawal}, {Paraskeva}, {Pursiainen}, {Raki{\'c}}, {Schulze}, {Taggart}, {Wedderkopp}, {Anderson}, {de Boer}, {Chambers}, {Chen}, {Damljanovi{\'c}}, {Fraser}, {Gao}, {Gomboc}, {Gromadzki}, {Ihanec}, {Maguire}, {Mar{\v{c}}un}, {M{\"u}ller-Bravo}, {Nicholl}, {Onori}, {Reynolds}, {Smartt}, {Sollerman}, {Smith}, {Wevers}, \& {Wyrzykowski}}]{Petrushevska2023}
{Petrushevska}, T., {Leloudas}, G., {Ili{\'c}}, D., {et~al.} 2023, \aap, 669, A140, \dodoi{10.1051/0004-6361/202244623}

\bibitem[{{Rees}(1988)}]{Rees1988}
{Rees}, M.~J. 1988, \nat, 333, 523, \dodoi{10.1038/333523a0}

\bibitem[{{Reynolds} {et~al.}(2022){Reynolds}, {Mattila}, {Efstathiou}, {Kankare}, {Kool}, {Ryder}, {Pe{\~n}a-Mo{\~n}ino}, \& {P{\'e}rez-Torres}}]{Reynolds2022}
{Reynolds}, T.~M., {Mattila}, S., {Efstathiou}, A., {et~al.} 2022, \aap, 664, A158, \dodoi{10.1051/0004-6361/202243289}

\bibitem[{{Roming} {et~al.}(2005){Roming}, {Kennedy}, {Mason}, {Nousek}, {Ahr}, {Bingham}, {Broos}, {Carter}, {Hancock}, {Huckle}, {Hunsberger}, {Kawakami}, {Killough}, {Koch}, {McLelland}, {Smith}, {Smith}, {Soto}, {Boyd}, {Breeveld}, {Holland}, {Ivanushkina}, {Pryzby}, {Still}, \& {Stock}}]{Roming2005}
{Roming}, P. W.~A., {Kennedy}, T.~E., {Mason}, K.~O., {et~al.} 2005, \ssr, 120, 95, \dodoi{10.1007/s11214-005-5095-4}

\bibitem[{{Roseboom} {et~al.}(2013){Roseboom}, {Lawrence}, {Elvis}, {Petty}, {Shen}, \& {Hao}}]{Roseboom2013}
{Roseboom}, I.~G., {Lawrence}, A., {Elvis}, M., {et~al.} 2013, \mnras, 429, 1494, \dodoi{10.1093/mnras/sts441}

\bibitem[{{S{\'a}nchez-S{\'a}ez} {et~al.}(2024){S{\'a}nchez-S{\'a}ez}, {Hern{\'a}ndez-Garc{\'\i}a}, {Bernal}, {Bayo}, {Calistro Rivera}, {Bauer}, {Ricci}, {Merloni}, {Graham}, {Cartier}, {Ar{\'e}valo}, {Assef}, {Concas}, {Homan}, {Krumpe}, {Lira}, {Malyali}, {Mart{\'\i}nez-Aldama}, {Mu{\~n}oz Arancibia}, {Rau}, {Bruni}, {F{\"o}rster}, {Pavez-Herrera}, {Tub{\'\i}n-Arenas}, \& {Brightman}}]{SanchezSaezet2024}
{S{\'a}nchez-S{\'a}ez}, P., {Hern{\'a}ndez-Garc{\'\i}a}, L., {Bernal}, S., {et~al.} 2024, \aap, 688, A157, \dodoi{10.1051/0004-6361/202347957}

\bibitem[{{Schartmann} {et~al.}(2009){Schartmann}, {Meisenheimer}, {Klahr}, {Camenzind}, {Wolf}, \& {Henning}}]{Schartmann2009}
{Schartmann}, M., {Meisenheimer}, K., {Klahr}, H., {et~al.} 2009, \mnras, 393, 759, \dodoi{10.1111/j.1365-2966.2008.14220.x}

\bibitem[{{Short} {et~al.}(2023){Short}, {Lawrence}, {Nicholl}, {Ward}, {Reynolds}, {Mattila}, {Yin}, {Arcavi}, {Carnall}, {Charalampopoulos}, {Gromadzki}, {Jonker}, {Kim}, {Leloudas}, {Mandel}, {Onori}, {Pursiainen}, {Schulze}, {Villforth}, \& {Wevers}}]{Short2023}
{Short}, P., {Lawrence}, A., {Nicholl}, M., {et~al.} 2023, \mnras, 525, 1568, \dodoi{10.1093/mnras/stad2270}

\bibitem[{{Spergel} {et~al.}(2015){Spergel}, {Gehrels}, {Baltay}, {Bennett}, {Breckinridge}, {Donahue}, {Dressler}, {Gaudi}, {Greene}, {Guyon}, {Hirata}, {Kalirai}, {Kasdin}, {Macintosh}, {Moos}, {Perlmutter}, {Postman}, {Rauscher}, {Rhodes}, {Wang}, {Weinberg}, {Benford}, {Hudson}, {Jeong}, {Mellier}, {Traub}, {Yamada}, {Capak}, {Colbert}, {Masters}, {Penny}, {Savransky}, {Stern}, {Zimmerman}, {Barry}, {Bartusek}, {Carpenter}, {Cheng}, {Content}, {Dekens}, {Demers}, {Grady}, {Jackson}, {Kuan}, {Kruk}, {Melton}, {Nemati}, {Parvin}, {Poberezhskiy}, {Peddie}, {Ruffa}, {Wallace}, {Whipple}, {Wollack}, \& {Zhao}}]{Spergel2015}
{Spergel}, D., {Gehrels}, N., {Baltay}, C., {et~al.} 2015, arXiv e-prints, arXiv:1503.03757, \dodoi{10.48550/arXiv.1503.03757}

\bibitem[{{Stalevski} {et~al.}(2016){Stalevski}, {Ricci}, {Ueda}, {Lira}, {Fritz}, \& {Baes}}]{Stalevski2016}
{Stalevski}, M., {Ricci}, C., {Ueda}, Y., {et~al.} 2016, \mnras, 458, 2288, \dodoi{10.1093/mnras/stw444}

\bibitem[{{Suganuma} {et~al.}(2006){Suganuma}, {Yoshii}, {Kobayashi}, {Minezaki}, {Enya}, {Tomita}, {Aoki}, {Koshida}, \& {Peterson}}]{Suganuma2006}
{Suganuma}, M., {Yoshii}, Y., {Kobayashi}, Y., {et~al.} 2006, \apj, 639, 46, \dodoi{10.1086/499326}

\bibitem[{{Tuna} {et~al.}(2025){Tuna}, {Metzger}, {Jiang}, \& {White}}]{Tuna2025}
{Tuna}, S., {Metzger}, B.~D., {Jiang}, Y.-F., \& {White}, C.~J. 2025, arXiv e-prints, arXiv:2501.13157, \dodoi{10.48550/arXiv.2501.13157}

\bibitem[{{van Velzen} {et~al.}(2016){van Velzen}, {Mendez}, {Krolik}, \& {Gorjian}}]{vV2016}
{van Velzen}, S., {Mendez}, A.~J., {Krolik}, J.~H., \& {Gorjian}, V. 2016, \apj, 829, 19, \dodoi{10.3847/0004-637X/829/1/19}

\bibitem[{{Wada}(2012)}]{Wada2012}
{Wada}, K. 2012, \apj, 758, 66, \dodoi{10.1088/0004-637X/758/1/66}

\bibitem[{{Wada}(2015)}]{Wada2015}
---. 2015, \apj, 812, 82, \dodoi{10.1088/0004-637X/812/1/82}

\bibitem[{{Wada} \& {Norman}(2002)}]{Wada2002}
{Wada}, K., \& {Norman}, C.~A. 2002, \apjl, 566, L21, \dodoi{10.1086/339438}

\bibitem[{{Wang} {et~al.}(2024){Wang}, {Ma}, {Wu}, \& {Jiang}}]{Wang2024}
{Wang}, M., {Ma}, Y., {Wu}, Q., \& {Jiang}, N. 2024, \apj, 960, 69, \dodoi{10.3847/1538-4357/ad0bfb}

\bibitem[{{Wang} {et~al.}(2022){Wang}, {Jiang}, {Wang}, {Zhu}, {Dou}, {Lin}, {Sun}, {Liu}, \& {Sheng}}]{Wang2022}
{Wang}, Y., {Jiang}, N., {Wang}, T., {et~al.} 2022, \apjl, 930, L4, \dodoi{10.3847/2041-8213/ac6670}

\bibitem[{{Wevers} \& {French}(2024)}]{Wevers2024-TDE}
{Wevers}, T., \& {French}, K.~D. 2024, \apjl, 969, L17, \dodoi{10.3847/2041-8213/ad5725}

\bibitem[{{Wevers} {et~al.}(2022){Wevers}, {Pasham}, {Jalan}, {Rakshit}, \& {Arcodia}}]{Wevers2022}
{Wevers}, T., {Pasham}, D.~R., {Jalan}, P., {Rakshit}, S., \& {Arcodia}, R. 2022, \aap, 659, L2, \dodoi{10.1051/0004-6361/202243143}

\bibitem[{{Wevers} {et~al.}(2024){Wevers}, {French}, {Zabludoff}, {Fischer}, {Rowlands}, {Guolo}, {Dalla Barba}, {Arcodia}, {Berton}, {Bian}, {Linial}, {Miniutti}, \& {Pasham}}]{Wevers2024-QPE}
{Wevers}, T., {French}, K.~D., {Zabludoff}, A.~I., {et~al.} 2024, \apjl, 970, L23, \dodoi{10.3847/2041-8213/ad5f1b}

\bibitem[{{Williamson} {et~al.}(2019){Williamson}, {H{\"o}nig}, \& {Venanzi}}]{Williamson2019}
{Williamson}, D., {H{\"o}nig}, S., \& {Venanzi}, M. 2019, \apj, 876, 137, \dodoi{10.3847/1538-4357/ab17d5}

\bibitem[{{Wright} {et~al.}(2010){Wright}, {Eisenhardt}, {Mainzer}, {Ressler}, {Cutri}, {Jarrett}, {Kirkpatrick}, {Padgett}, {McMillan}, {Skrutskie}, {Stanford}, {Cohen}, {Walker}, {Mather}, {Leisawitz}, {Gautier}, {McLean}, {Benford}, {Lonsdale}, {Blain}, {Mendez}, {Irace}, {Duval}, {Liu}, {Royer}, {Heinrichsen}, {Howard}, {Shannon}, {Kendall}, {Walsh}, {Larsen}, {Cardon}, {Schick}, {Schwalm}, {Abid}, {Fabinsky}, {Naes}, \& {Tsai}}]{Wright2010}
{Wright}, E.~L., {Eisenhardt}, P. R.~M., {Mainzer}, A.~K., {et~al.} 2010, \aj, 140, 1868, \dodoi{10.1088/0004-6256/140/6/1868}

\bibitem[{{Xiong} {et~al.}(2025){Xiong}, {Jiang}, {Pan}, {Hao}, \& {Li}}]{Xiong2025}
{Xiong}, Y., {Jiang}, N., {Pan}, Z., {Hao}, L., \& {Li}, Z. 2025, arXiv e-prints, arXiv:2503.19722.
\newblock \doarXiv{2503.19722}

\bibitem[{{Yao} {et~al.}(2023){Yao}, {Ravi}, {Gezari}, {van Velzen}, {Lu}, {Schulze}, {Somalwar}, {Kulkarni}, {Hammerstein}, {Nicholl}, {Graham}, {Perley}, {Cenko}, {Stein}, {Ricarte}, {Chadayammuri}, {Quataert}, {Bellm}, {Bloom}, {Dekany}, {Drake}, {Groom}, {Mahabal}, {Prince}, {Riddle}, {Rusholme}, {Sharma}, {Sollerman}, \& {Yan}}]{Yao2023}
{Yao}, Y., {Ravi}, V., {Gezari}, S., {et~al.} 2023, \apjl, 955, L6, \dodoi{10.3847/2041-8213/acf216}

\bibitem[{{Yuan} {et~al.}(2024){Yuan}, {Winter}, \& {Lunardini}}]{Yuan2024}
{Yuan}, C., {Winter}, W., \& {Lunardini}, C. 2024, \apj, 969, 136, \dodoi{10.3847/1538-4357/ad50a9}

\bibitem[{Zhou {et~al.}(2024{\natexlab{a}})Zhou, Huang, Guo, Li, \& Pan}]{Zhou2024a}
Zhou, C., Huang, L., Guo, K., Li, Y.-P., \& Pan, Z. 2024{\natexlab{a}}, Phys. Rev. D, 109, 103031, \dodoi{10.1103/PhysRevD.109.103031}

\bibitem[{{Zhou} {et~al.}(2025){Zhou}, {Zeng}, \& {Pan}}]{Zhou2025}
{Zhou}, C., {Zeng}, Y., \& {Pan}, Z. 2025, \apj, 985, 242, \dodoi{10.3847/1538-4357/adcee2}

\bibitem[{Zhou {et~al.}(2024{\natexlab{b}})Zhou, Zhong, Zeng, Huang, \& Pan}]{Zhou2024b}
Zhou, C., Zhong, B., Zeng, Y., Huang, L., \& Pan, Z. 2024{\natexlab{b}}, Phys. Rev. D, 110, 083019, \dodoi{10.1103/PhysRevD.110.083019}

\end{thebibliography}

\end{document}